\documentclass[twocolumn,twocolappendix]{aastex63}

\usepackage{newtxtext,newtxmath}

\usepackage[T1]{fontenc}

\DeclareRobustCommand{\VAN}[3]{#2}
\let\VANthebibliography\thebibliography
\def\thebibliography{\DeclareRobustCommand{\VAN}[3]{##3}\VANthebibliography}

\usepackage{graphicx}	
\usepackage{amsmath}	

\usepackage{amssymb}	
\usepackage{multirow}
\usepackage{indentfirst}
\usepackage{lineno}

\usepackage{url}

\newcommand{\be}{\begin{equation}}
\newcommand{\ee}{\end{equation}}
\newcommand{\ba}{\begin{eqnarray}}
\newcommand{\ea}{\end{eqnarray}}
\newcommand{\bi}{\begin{itemize}}
\newcommand{\ei}{\end{itemize}}

\usepackage{color}

\newcommand{\add}[1]{\textcolor{black}{#1}}

\received{XXX}
\revised{YYY}
\accepted{ZZZ}

\submitjournal{ApJ}

\shorttitle{tSZ effect}
\shortauthors{Chen et al.}

\begin{document}

\title{Thermal energy census with the Sunyaev-Zel'dovich effect of DESI galaxy clusters/groups and its implication on the weak lensing power spectrum }

\correspondingauthor{Ziyang Chen,Pengjie Zhang}
\email{chen\_zy@sjtu.edu.cn,zhangpj@sjtu.edu.cn}

\author[0000-0001-8648-1789]{Ziyang Chen}
\affiliation{Department of Astronomy, School of Physics and Astronomy, Shanghai Jiao Tong University, Shanghai, 200240, China}
\affiliation{Key Laboratory for Particle Astrophysics and Cosmology
(MOE)/Shanghai Key Laboratory for Particle Physics and Cosmology,China}

\author{Pengjie Zhang}
\affiliation{Department of Astronomy, School of Physics and Astronomy, Shanghai Jiao Tong University, Shanghai, 200240, China}
\affiliation{Tsung-Dao Lee Institute, Shanghai Jiao Tong University , Shanghai 200240, China}
\affiliation{Key Laboratory for Particle Astrophysics and Cosmology
(MOE)/Shanghai Key Laboratory for Particle Physics and Cosmology,China}

\author{Xiaohu Yang}
\affiliation{Department of Astronomy, School of Physics and Astronomy, Shanghai Jiao Tong University, Shanghai, 200240, China}
\affiliation{Tsung-Dao Lee Institute, Shanghai Jiao Tong University , Shanghai 200240, China}
\affiliation{Key Laboratory for Particle Astrophysics and Cosmology
(MOE)/Shanghai Key Laboratory for Particle Physics and Cosmology,China}

\begin{abstract}

	We carry out a thermal energy census of hot baryons at $z < 1$, by cross-correlating the \emph{Planck} MILCA y-map with 0.8 million clusters/groups selected from the Yang et.al (2021) catalog. The thermal Sunyaev-Zel'dovich (tSZ) effect around these clusters/groups are reliably obtained, which enables us to make our model constraints based on one-halo (1h) and two-halo (2h) contributions, respectively. (1) The total measurement S/N of the one-halo term is 63. We constrain the $Y$-$M$ relation over the halo mass range of $10^{13}$-$10^{15} M_\odot/h$, and find $Y\propto M^{\alpha}$ with $\alpha= 1.8$ at $z=0.14$ ($\alpha=2.1$ at $z=0.75$). The total thermal energy of gas bound to clusters/groups increases from $0.1\ \rm meV/cm^3$ at $z=0.14$ to $0.22\ \rm meV/cm^3$ at $z=0.75$. (2) The two-halo term is used to constrain the bias-weighted electron pressure $\langle b_yP_e \rangle$. We find that $\langle b_yP_e \rangle$ (in unit of $\rm meV/cm^3$) increases from $0.24\pm 0.02$ at $z=0.14$ to $0.45\pm 0.02$ at $z=0.75$. These results lead to several implications. (i) The hot gas fraction $f_{\rm gas}$ in clusters/groups monotonically increase with halo mass, where $f_{\rm gas}$ of a $10^{14} M_\odot/h$ halo is $\sim 50\%$ ($25\%$) of the cosmic mean at $z=0.14\ (0.75)$. (ii) By comparing the 1h- and 2h-terms, we obtain tentative constraint on the thermal energy of unbound gas. (iii) The above results lead to significant suppression of matter and weak lensing power spectrum at small scales. These implications are important for astrophysics and cosmology, and we will further investigate them with improved data  and gas modeling.
\end{abstract}


\section{Introduction} \label{sec:intro}

Hot, free electrons in the late universe scatter off CMB photons through the inverse Compton scattering and generate the secondary CMB anisotropies. This is the famous thermal Sunyaev-Zel'dovich (tSZ) effect \citep{Sunyaev1972, Carlstrom2002, Kitayama2014}. It induces a temperature fluctuation $\Delta T_{\rm tSZ}$ with a characteristic spectral dependence ($g(x)$) and amplitude described by the Compton y-parameter:
	\ba
		\frac{\Delta T_{\rm tSZ}}{T_{\rm CMB}}=g(x)y\ ,\ y=\frac{\sigma_T}{m_ec^2}\int n_ek_B T ad\chi\,,
	\ea
	where $g(x)=x \coth(x /2) -4$ and $x \equiv {h\nu}/k_{\rm B}T_{\rm CMB}$. 

    The tSZ effect contains important information regarding the astrophysics of clusters/groups and cosmology. In the first, the tSZ effect is a direct probe of cluster pressure profile \citep{Ruppin2018, Pandey2021, Ma2021}, baryon abundance \citep{HM2006, Lim2018} and distribution \citep{LeBrun2015, Ma2015, Kim2021, Meinke2021, Amodeo2021, Kim2022}. Even the baryons in filaments can be inferred by stacking tSZ maps of clusters pairs \citep{Munoz2018, Graaff2019, Tanimura2019, Gouin2022}.
    Also, it can shed light on constraining the strength of supernova and AGN feedback \citep{Hojjati2017, Spacek2018, Troster2021, Gatti2021, chen2023}.
	In addition, the cross-correlation of tSZ with galaxy distribution \citep{Zhang2001, Hill2018, Pandey2020} or a weak lensing survey  \citep{Shao2011, Hojjati2015, Gatti2022, Pandey2022} can increase the measurement significance of tSZ significantly. These kinds of cross-correlation have enabled the measurement of the mean bias-weighted pressure $\langle b_yP_e \rangle$ as a function of redshift \citep{VanWaerbeke2014, Vikram2017, Koukou2020, Chiang2020, Yan2021}. 
	Furthermore, the tSZ effect can be also used to constrain cosmological parameters such as $\sigma_8$ \citep{Komatsu2002,Zhang02,Horowitz2017, Osato2020}, dark energy properties \citep{Bolliet2018} and the evolution of $T_{\rm CMB}$ \citep{Hurier2014}.
	
	Direct measurement of cluster tSZ effect by \emph{Planck}, ACT and SPT is limited to clusters with mass $\ga 2\times 10^{14} M_\odot/h$ \citep{ Planck2014_SZ_catalog,Marriage2011, Hasselfield2013,Reichardt2013, Bleem2015,Brodwin2015}. Since a noticeable fraction of thermal energy comes from less massive clusters/groups, the above measurements are incapable of carrying out a complete thermal energy census of clusters/groups. On the other hand, the cross-correlation measurement with galaxies only measures $\langle b_yP_e \rangle$ and lacks detailed information on the thermal energy distribution. 
	The recently released group catalog \citep{Yang2021} provides us a good opportunity to contrain both the thermal energy and $\langle b_yP_e \rangle$ as a function of halo mass down to $10^{13} M_\odot/h$. This group catalog contains about a million clusters/groups robustly identified (with richness $\geq 5$) in the $z<1$ universe. This data set is not only large in cluster number, but also has reasonable completeness and redshift/mass estimation. It has enabled us to measure the CMB lensing with $S/N\simeq 40$ \citep{Sun22}, and the kinematic Sunyaev Zel'dovich (kSZ) effect with $S/N\simeq 5$ \citep{Chen22}. Given that cluster/group tSZ is significantly  stronger than kSZ, we expect high S/N in the tSZ measurement. 
	This measurement will provide valuable information on the thermal energy distribution in the universe, and shed light on important gastrophysics such as feedback. It can also put useful constraint on the baryonic effect on weak lensing cosmology. 

	This paper is organized as follows. We first introduce the data in \S\ref{sec:data}, and then present the method of measuring tSZ in \S\ref{sec:method}. The results are shown and analyzed in \S\ref{sec:result}. We show the implications of this measurement in \S\ref{sec:appli} and finally present our conclusions and discussions in \S\ref{sec:conclu}. We also include an appendix to explain further details and tests. we adopt a flat cosmology with parameters: $h=0.676$, $\Omega_{\rm dm}h^2=0.119$, $\Omega_b h^2=0.022$, $\sigma_8=0.81$ and $n_s=0.967$ \citep{Planck2020}.

\section{Data}\label{sec:data}
\subsection{Planck Compton parameter map}

The \emph{Planck} collaboration released the full-sky Compton parameter map (y-map) of tSZ constructed by two algorithms, NILC and MILCA \citep{Planck2016}\footnote{Based on observations obtained with Planck (http://www.esa.int/Planck), an ESA science mission with instruments and contributions directly funded by ESA Member States, NASA, and Canada.}. These two methods are both based on the Internal Linear Combination (ILC) method and the known spectrum of CMB components. The difference is the method to calculate the optimal scale-independent and spatially-varying linear weight. The performance of NILC and MILCA do not show distinguishable differences in many studies \citep{Vikram2017, Koukou2020}. Besides, a higher noise level is shown in the large-scale of NILC map. Therefore, we choose to utilize the MILCA map to measure y-profile of clusters. This map has a circular Gaussian beam of 10 arcmin and nside=2048 for healpix pixelization resolution \citep{Planck2016}.
	
To reduce the contamination from residual Galactic foregrounds, we apply a combination of Planck Galactic mask with 40\% sky coverage and a point source mask from the foreground masks used for the Compton parameter analysis provided by \emph{Planck}.

\subsection{DESI group catalog}

In this work, we use the DESI group catalog obtained by \citet{Yang2021} \add{(Y21)} from the Data Release 9 (DR9) of the DESI Legacy Imaging survey. This group catalog was constructed using an extended version of the halo-based group finder developed by \citet{Yang2005, Yang2007}, which can use photometric or spectroscopic redshifts simultaneously for galaxies. The biggest advantage of this catalog for our concern is that \add{the global completeness and overall purity of the detected groups is high. The completeness and purity of clusters with mass larger than $10^{14}M_{\odot}/h$ is close to one and the completeness of groups with mass $>10^{12}$ is from 70\% to 80\%.} It is well-known that most of the thermal energy of the universe resides in these massive clusters. The tSZ effect, which is proportional to the thermal energy of baryon, can be detected with high measurement significance with this catalog. \add{And the large sample size also enables the detection of thermal contribution from the small halos.} In addition, this catalog provides an reliable estimation of the cluster redshift and mass. The redshift accuracy for groups with more than 10 members is about 0.008. \add{The dark matter halos are defined having an overdensity of 180 times larger than the mean universe background density.} And the uncertainty of halo mass is about 0.2 dex for the massive clusters ($> 10^{13.5} M_\odot/h$) and about 0.40 dex at the low-mass end ($\sim 10^{12} M_\odot/h$).
	
When measuring the tSZ effect in this work, we only use the clusters with at least 5 members. Because there are relatively large uncertainties in mass and redshift estimation for clusters with $N_g < 5$. And these uncertainties would bias the measurement in an unexpected way. In addition, the clusters with small richness are usually small-mass halos or have a higher probability to be misidentified. Moreover, as we have tested,  including them in our measurements would not improve the S/N significantly.

\begin{deluxetable*}{cccccccccccc}
\tablecaption{The detailed information of tSZ stacking measurement and fitting in different mass and redshift bins. \label{tab:all}}
	\tablehead{\colhead{range of $z$} & \colhead{range of $\lg M_{\rm L}$} & \colhead{$N_{\rm cluster}$} & \colhead{$\lg M_t$} & \colhead{$\bar b_g$} & \colhead{$A_1$} & \colhead{S/N($A_1$)} & \colhead{$A_2$} & \colhead{S/N($A_2$)} & \colhead{$A_3\times 10^8$} & \colhead{S/N($A_3$)} & \colhead{$\chi^2_{\rm min}$} 
	}
	\startdata
	$[$0.0, 0.2) & $[$14.7, 15.5) & 215 & 14.71 & 4.44 & 0.647 & 7.67 & -2.53 & 0.8 & 27.47 & 1.2 & 1.782 \\
    $[$0.0, 0.2) & $[$14.5, 14.7) & 450 & 14.47 & 3.36 & 0.631 & 10.56 & 2.44 & 2.5 & -8.97 & 1.6 & 0.394 \\
    $[$0.0, 0.2) & $[$14.3, 14.5) & 951 & 14.32 & 2.92 & 0.487 & 11.53 & 1.19 & 2.5 & -3.09 & 1.1 & 1.588 \\
    $[$0.0, 0.2) & $[$14.1, 14.3) & 1932 & 14.12 & 2.56 & 0.451 & 9.22 & 1.11 & 4.4 & -3.89 & 2.2 & 4.968 \\
    $[$0.0, 0.2) & $[$13.9, 14.1) & 3468 & 13.92 & 2.18 & 0.495 & 7.69 & 0.74 & 3.0 & -1.39 & 1.0 & 2.504 \\
    $[$0.0, 0.2) & $[$13.5, 13.9) & 17054 & 13.6 & 1.79 & 0.455 & 8.72 & 0.44 & 5.9 & -0.66 & 1.3 & 3.216 \\
    $[$0.0, 0.2) & $[$13, 13.5) & 60269 & 13.17 & 1.4 & 0.364 & 2.22 & 0.28 & 5.8 & -0.18 & 0.5 & 5.949 \\
    \hline
    $[$0.2, 0.4) & $[$14.9, 15.5) & 280 & 14.9 & 5.99 & 0.439 & 10.96 & -0.89 & 0.8 & 2.42 & 0.5 & 12.94 \\
    $[$0.2, 0.4) & $[$14.8, 14.9) & 302 & 14.71 & 4.92 & 0.511 & 12.41 & 0.68 & 0.6 & 2.51 & 0.7 & 4.807 \\
    $[$0.2, 0.4) & $[$14.7, 14.8) & 573 & 14.61 & 4.4 & 0.499 & 12.02 & 0.94 & 1.5 & -0.24 & 0.1 & 2.702 \\
    $[$0.2, 0.4) & $[$14.6, 14.7) & 920 & 14.53 & 3.98 & 0.476 & 14.04 & 1.83 & 3.1 & -4.57 & 2.4 & 9.233 \\
    $[$0.2, 0.4) & $[$14.5, 14.6) & 1467 & 14.43 & 3.65 & 0.424 & 15.88 & 1.3 & 3.4 & -2.69 & 2.0 & 8.437 \\
    $[$0.2, 0.4) & $[$14.4, 14.5) & 2463 & 14.38 & 3.39 & 0.387 & 16.22 & 0.83 & 2.5 & 0.81 & 0.7 & 7.384 \\
    $[$0.2, 0.4) & $[$14.3, 14.4) & 3728 & 14.28 & 3.21 & 0.33 & 12.74 & 1.08 & 4.1 & -0.7 & 0.8 & 4.215 \\
    $[$0.2, 0.4) & $[$14.2, 14.3) & 5541 & 14.18 & 3.01 & 0.322 & 10.08 & 0.65 & 3.3 & -0.24 & 0.4 & 6.999 \\
    $[$0.2, 0.4) & $[$14, 14.2) & 19693 & 14.02 & 2.65 & 0.325 & 11.98 & 0.63 & 5.2 & 0.18 & 0.5 & 1.189 \\
    $[$0.2, 0.4) & $[$13.7, 14) & 67347 & 13.76 & 2.19 & 0.27 & 7.7 & 0.54 & 9.9 & -0.18 & 0.9 & 6.463 \\
    $[$0.2, 0.4) & $[$13.4, 13.7) & 145052 & 13.46 & 1.81 & 0.249 & 3.72 & 0.47 & 11.2 & -0.06 & 0.4 & 1.983 \\
    $[$0.2, 0.4) & $[$13, 13.4) & 252970 & 13.15 & 1.51 & 0.382 & 2.37 & 0.45 & 15.1 & -0.0 & 0.0 & 9.379 \\
    \hline
    $[$0.4, 0.6) & $[$14.9, 15.5) & 205 & 14.89 & 6.91 & 0.389 & 9.45 & 2.98 & 1.5 & 1.13 & 0.4 & 9.444 \\
    $[$0.4, 0.6) & $[$14.8, 14.9) & 208 & 14.71 & 5.75 & 0.446 & 10.3 & 5.17 & 2.3 & -2.52 & 0.7 & 10.0 \\
    $[$0.4, 0.6) & $[$14.7, 14.8) & 483 & 14.61 & 5.15 & 0.517 & 14.18 & 1.33 & 1.1 & 0.47 & 0.2 & 3.078 \\
    $[$0.4, 0.6) & $[$14.6, 14.7) & 961 & 14.53 & 4.64 & 0.456 & 13.24 & 2.64 & 3.1 & -0.8 & 0.4 & 6.109 \\
    $[$0.4, 0.6) & $[$14.5, 14.6) & 1780 & 14.43 & 4.24 & 0.457 & 15.28 & 1.76 & 2.8 & 0.5 & 0.4 & 3.042 \\
    $[$0.4, 0.6) & $[$14.4, 14.5) & 2885 & 14.37 & 3.95 & 0.334 & 11.64 & 2.51 & 4.6 & -0.14 & 0.2 & 2.592 \\
    $[$0.4, 0.6) & $[$14.3, 14.4) & 4846 & 14.28 & 3.74 & 0.342 & 10.64 & 1.28 & 3.5 & 1.21 & 1.7 & 3.771 \\
    $[$0.4, 0.6) & $[$14.2, 14.3) & 7644 & 14.18 & 3.49 & 0.315 & 9.5 & 1.49 & 4.9 & 0.15 & 0.3 & 7.484 \\
    $[$0.4, 0.6) & $[$14, 14.2) & 28703 & 14.02 & 3.06 & 0.273 & 9.71 & 1.29 & 8.8 & 0.13 & 0.5 & 12.015 \\
    $[$0.4, 0.6) & $[$13.7, 14) & 104894 & 13.76 & 2.51 & 0.218 & 5.53 & 0.88 & 10.5 & 0.44 & 3.2 & 10.758 \\
    $[$0.4, 0.6) & $[$13.4, 13.7) & 218311 & 13.47 & 2.07 & 0.161 & 2.34 & 0.59 & 11.7 & 0.6 & 5.9 & 9.083 \\
    $[$0.4, 0.6) & $[$13, 13.4) & 225638 & 13.17 & 1.7 & 0.399 & 1.88 & 0.54 & 10.9 & 0.81 & 7.8 & 5.744 \\
    \hline
    $[$0.6, 1.0) & $[$14.7, 15.5) & 465 & 14.67 & 6.79 & 0.441 & 11.94 & 3.39 & 1.6 & 0.78 & 0.4 & 2.643 \\
    $[$0.6, 1.0) & $[$14.6, 14.7) & 752 & 14.53 & 5.63 & 0.356 & 8.49 & 4.44 & 2.5 & -3.27 & 1.8 & 4.735 \\
    $[$0.6, 1.0) & $[$14.5, 14.6) & 1674 & 14.43 & 5.17 & 0.343 & 9.6 & 2.85 & 2.7 & 0.5 & 0.5 & 3.537 \\
    $[$0.6, 1.0) & $[$14.4, 14.5) & 3205 & 14.37 & 4.82 & 0.286 & 10.55 & 3.51 & 4.6 & 0.51 & 0.6 & 1.61 \\
    $[$0.6, 1.0) & $[$14.3, 14.4) & 6196 & 14.27 & 4.57 & 0.285 & 8.14 & 2.58 & 4.5 & 0.44 & 0.8 & 4.07 \\
    $[$0.6, 1.0) & $[$14.2, 14.3) & 10698 & 14.17 & 4.27 & 0.264 & 7.76 & 2.37 & 5.2 & 0.14 & 0.3 & 4.865 \\
    $[$0.6, 1.0) & $[$14.1, 14.2) & 17532 & 14.08 & 3.89 & 0.263 & 7.16 & 1.94 & 5.8 & 0.83 & 2.3 & 6.582 \\
    $[$0.6, 1.0) & $[$13.5, 14.1) & 187974 & 13.76 & 3.0 & 0.182 & 5.88 & 1.22 & 14.9 & 0.41 & 4.4 & 8.179 \\
    \hline
	\enddata
	\tablecomments{The first and second columns are the redshift and the luminosity mass range of clusters in each bin. The third column is the number of clusters. 
	The fourth and fifth column is the mean calibrated mass (details in Appendix.\ref{app:mtml}) and cluster bias. The 6th and 7th columns are the best-fitted value and S/N of the coefficient for the 1h-term. The 8th to 11th columns are those for the 2h-term and the background term. The last column is the $\chi^2_{\rm min}$ for fitting in this bin.
	The number of data points in each bin is 40. But due to strong correlation between adjacent data points, the number of independent data points is $\sim 10$. Given 3 fitting parameters, $\chi^2_{\rm min}\la 10$ is reasonable. Groups with $M_{\rm L} < 10^{13} M_\odot/h$ ($M_{\rm L} < 10^{13.5} M_\odot/h$ for $0.6\leq z \leq 1$) are not used to measure the $Y-M$, $f_{\rm gas}-M$ relation and $\langle b_yP_e \rangle$ due to relatively high $\chi^2_{\rm min}$ and significant detection of the background term (shown in Appendix. \ref{APP:low_mass_sample}).}
\end{deluxetable*}

To investigate how the tSZ effect may depend on the cluster mass and redshift, we divide the cluster sample into several redshift and mass bins. First, we separate clusters/groups in our sample into  4 redshift bins: $0.\leq z< 0.2$, $0.2\leq z < 0.4$, $0.4\leq z < 0.6$ and $0.6\leq z \leq 1$. Then, the binning for cluster mass is applied so that we can obtain reliable tSZ effect measurements.  In total we have 39 cluster/group subsamples for our subsequent study. The details of selecting the clusters/groups in different redshift and halo mass bins, as well as their numbers, are outlined in the first three columns of Table. \ref{tab:all}.

\section{Method}\label{sec:method}

To obtain a high significance measurement of the tSZ effect, the secondary anisotropy of CMB, we stack the \emph{Planck} MILCA y-map at the position of galaxy clusters.  From DESI group catalog DR9, we can obtain the coordinates, i.e., Ra, Dec and redshift, of clusters. Then the tSZ plane surrounding each cluster would be cut from the MILCA y-map with the flat approximation. The length of the plane is set to be 160 arcmin and divided into $101\times 101$ grids. The cluster is positioned at the center (origin point) of the plane. The value of a grid point is set to be the value of the pixel which it locates in. Here, we assume the y-profile of a cluster is circular symmetric. So the direction of x, y-axis on the stacking plane can be chosen randomly. Finally, the stacked y-profile around clusters is
\ba
		\hat y(\theta_i)=\frac{\sum_j w_j y_j(\vec{\theta})}{\sum_j w_j},
\ea
where $j$ represents the $j$-th cluster in a sample and the length of $\vec{\theta}$ is belong to the $i$-th $\theta$-bin. To avoid contamination from Galactic foreground, we adopt the combination of a 40\% Galactic Mask and a point source mask both provided by \emph{Planck}. The pixel within the masked region has the weight $w_j=0$, otherwise $w_j=1$.
	
The stacked y-profile $\hat y(\theta)$ is the combination of a 1h-term, a 2h-term and a background term coming from residual CMB or other components. Below, we would present how we obtain the template of 1h- and 2h-terms and how to fit the coefficient of each term.

\subsection{1h-term profile} \label{subsec:M_1h}

	To model the 1h-term of $y$-profile, we adopt the Komatsu-Seljak (KS) gas density and temperature profiles \citep{Komatsu2001} as the baseline. Note here we only keep the shapes of the pressure profile fixed as the KS prediction, but treat its amplitude as a free parameter. This parameter is a major indicator of cluster gas fraction and intracluster gastrophysics. Gastrophysics may also alter the halo concentration and therefore change the pressure profile. Since the Planck angular resolution does not allow us to put useful constraint on its variation, we will only discuss its impact in Appendix. \ref{app:concentration}. 
	
	The KS profile assumes the gas is in hydrostatic equilibrium and the gas density profile tracks the DM profiles in the outer region of a halo. Here, we follow the simplified version in \citet{Martizzi2013, Mead2020} with a fixed polytropic index $\Gamma$. The model is described in below briefly. The density profile of gravitationally bound gas is 
	\ba
		\rho_{\rm gas}(M,r) \propto \left[\frac{ln(1+r/r_s)}{r/r_s}\right]^{1/(\Gamma -1)},
	\ea
	where $\Gamma=1.17$ and $r_s=r_\nu/c$ is the ratio of virial radius and concentration. The concentration-mass relation is 
	\ba
		c(M)=7.85(\frac{M}{2*10^{12}h^{-1}M_\odot})^{-0.081}(1+z)^{-0.71}
	\ea
	\citep{Duffy2008}. Further analysis may adopt more accurate $c$-$M$ relation such as that of \citet{Zhao2009}, in particular at higher redshift. 
	The normalization of the above profile is 
	\ba
		f_{\rm gas}(M)M=\int^{r_v}_04\pi r^2 \rho_{\rm gas}(M,r)dr.
	\ea
	It is expected that  $f_{\rm gas} = {\Omega_b}/{\Omega_m}$ for sufficiently massive clusters, while feedback will reduce its value. This $f_{\rm gas}$ (equivalently the $A_1$ fitting parameter which will be introduced later) is a major parameter that our measurement will constrain. 
	
	The temperature profile is fixed by the hydrostatic equilibrium 
	\ba
	\label{eq:Tg}
		T_g(M,r)=T_v(M)\frac{\ln (1+r/r_s)}{r/r_s},
	\ea
	where $T_v(M)$ is the virial temperature
	\begin{eqnarray}
		\frac 32 k_BT_v(M) = \frac{GMm_p\mu_p}{ar_v}.
	\end{eqnarray}
	$m_p$ is the proton mass, $\mu_p$ is the mean gas particle mass divided by the proton mass, $r_v$ is the comoving virial radius and $a$ is the scale factor. The electron pressure is the product of density and temperature profile
	\ba\label{eq:pe}
		P_e(M,r)=\frac{\rho_{\rm gas}(M,r)}{m_p\mu_e}k_BT_g(M,r),
	\ea
	where $\mu_e=1.17$ is the mean gas particle mass divided by the proton mass. In the last, the 1h-term y-profile is the integration of $P_e(M,r)$ along the line-of-sight
	\ba\label{eq:y}
		y(\theta)=\frac{\sigma_T}{m_ec^2}\int \frac{d\chi}{1+z}P_e(\theta|\chi).
	\ea
	
	The beam size of \emph{Planck} has also been taken into account and approximated using a Gaussian function $W(l)=\exp(-{(l \sigma_{\rm beam})^2}/{2})$. For \emph{Planck} MILCA map, $\sigma_{\rm beam}={\rm FWHM}/{2\sqrt{2\ln 2}}=4.25\ \rm arcmin$. Therefore, the final 1h-term profile is 
	\ba \label{eq:y1}
		y_1(\theta) = \int \frac{ldl}{2\pi} J_0(\theta l)y(l)W(l).
	\ea
	$y(l)$ is the Hankel transformation of $y(\theta)$
	\ba 
		y(l)=2\pi\int J_0(l\theta )y(\theta)\theta d\theta
	\ea
	In addition, the mis-centering effect in cluster determination cause a similar effect as the beam. So we replace $\sigma_{\rm beam}$ as $\sigma_{\rm eff}\equiv \sqrt{\sigma_{\rm beam}^2+\sigma_{\rm mc}^2}$ to take mis-centering into account
	\ba \label{eq:mc}
		\sigma_{\rm mc}=\eta_{\rm mc}\frac{r_\nu}{d_c},
	\ea
	$r_\nu$ and $d_c$ are the virial radius and comoving distance of a cluster. The default value of $\eta_{\rm mc}$ is set to 0.2. We will discuss how this parameter would influence the results in Appendix. \ref{app:miscentering}.

\subsection{2h-term profile}
	
	The 2h-term would also contribute to the stacked y-profile $\hat y(\theta)$. Its profile is 
	\ba
		\int \bar{P_e}\xi_{g,p}\frac{\sigma_T a d\chi}{m_ec^2} \propto \\\nonumber &&\frac{\sigma_T}{(1+z)m_ec^2}\int\xi_{gm}(\theta,r_{\parallel}|z_g)dr_{\parallel}.
	\ea
	Here, we assume on large scale the gas distribution follows that of DM. The integration of the correlation function is 
	\ba
		\int\xi_{gm}(r_{\perp},r_{\parallel}|z_g)&& dr_{\parallel}= \\\nonumber &&
		b_g\int P_m(k_\perp, z)e^{ik_\perp r_\perp \cos\phi}\frac{k_\perp dk_\perp d\phi}{(2\pi)^2},
	\ea
    where $b_g$ is the bias of the clusters/groups in consideration, $P_m(k,z)$ is the matter power spectrum at redshift $z$. We estimate $ b_g $ from cluster mass distribution and the bias-mass relation \citep{sheth2001}. Since
	\ba
		\int^{2\pi}_0 e^{ik_\perp r_\perp \cos\phi}=2\pi J_0(k_\perp r_\perp),
	\ea
	and the template of 2h-term profile is 
	
	\ba \label{eq:y2}
		y_2(\theta)&=& b_g \int P_m(k_\perp, z)e^{-\frac{k^2_\perp\sigma^2_{\rm beam}d^2_c(z)}{2}}J_0(k_\perp r_\perp) \frac{k_\perp d k_\perp}{2\pi}.
	\ea
	Here, we consider the smoothing effect from beam size of CMB survey.

\begin{figure*}
		\centering
 		\includegraphics[width=0.95\textwidth, height=0.6\textwidth]{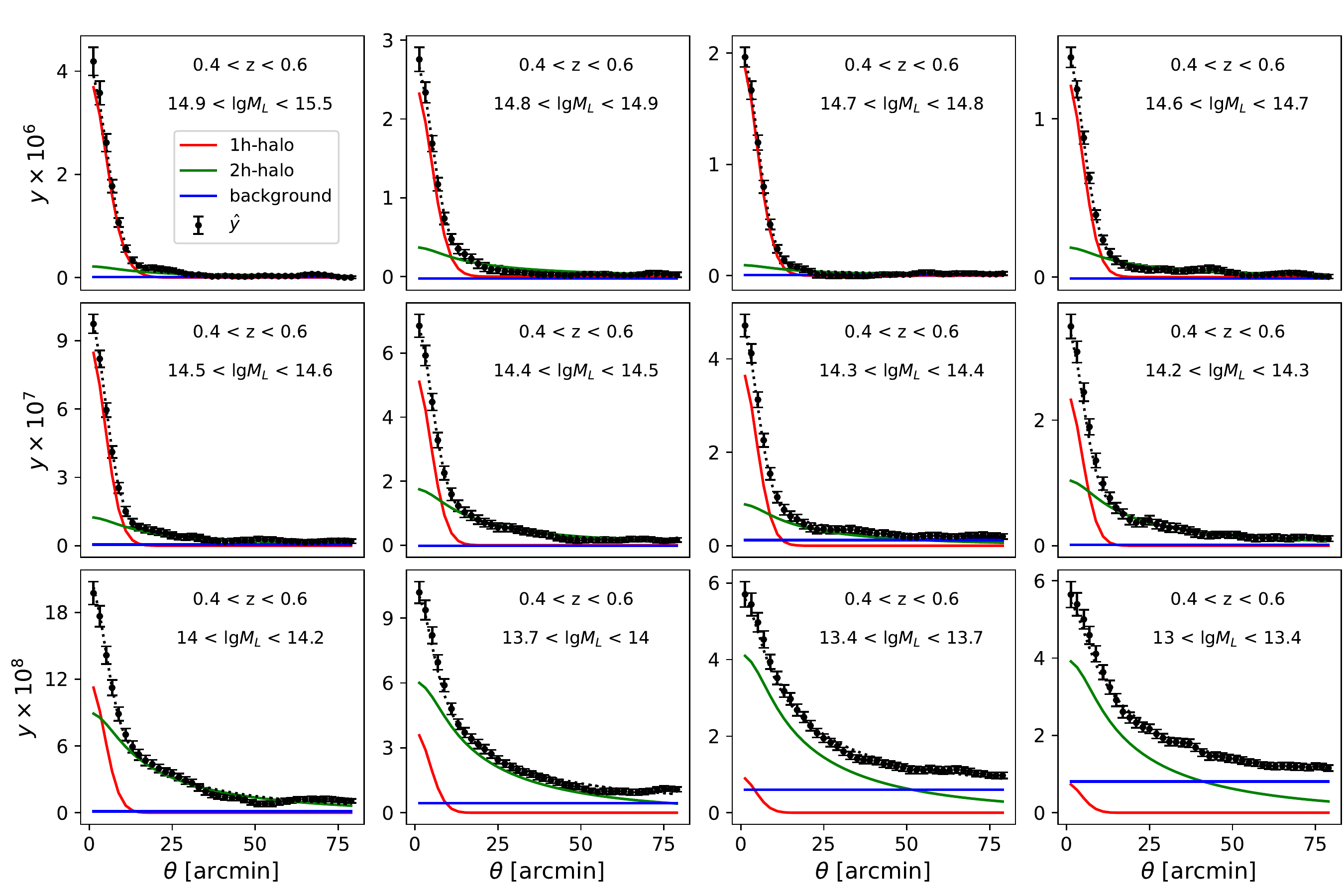}
 		\caption{The stacking results of tSZ measurements as a function of angular distance to the center of clusters. Each panel shows the result of a mass bin with $0.4 \leq z < 0.6$. The black dots are the stacking results with errorbars estimated by Jackknife resampling. The red, green and blue lines represent the bestfit one-halo, 2h-and the background term. The dotted lines are the sum of these three terms.\label{fig:y_profile}}
\end{figure*}

\subsection{Fitting}

	We assume the measured stacking tSZ profile contains three components and the theoretical model is
	\ba
		y^{\rm th} (\theta)=A_1 y_1(\theta)+ A_2 y_2(\theta) +A_3 y_3(\theta).
	\ea
	$A_1$ is the coefficient of the 1h-term profile within our methodology. Its physical meaning is the ratio of the fraction of gas in a cluster and the mean fraction of baryon in the universe. 
	\ba
		A_1 = \frac{f_{\rm gas}}{\Omega_b/\Omega_m}.
	\ea 
	This free parameter captures how much baryons is blown away by feedback processes. $A_2$ is the coefficient of the 2h-term 
	\ba
		A_2 = \langle  b_yP_e \rangle.
	\ea
	 In addition, $y_3(\theta) = \textbf{1}$ represents a scale-independent background term, in order to consider residuals of other CMB components which could contaminate the MILCA y-map.
	
	From Eq.\ref{eq:y1} and \ref{eq:y2}, the 1h-term template $y_1(\theta)$ relies on the mass and redshift distribution of the stacked cluster sample, while the 2h-template $y_2(\theta)$ only relies on the redshift distribution.
	It is worth noting that the mass estimated in DESI group catalog DR9 is higher than the true value and the uncertainty is about 0.2 to 0.4 dex from high-mass to low-mass end (as shown in Fig.9 in \citet{Yang2021}). We discuss in Appendix. \ref{app:mtml} about how to obtain an unbiased 1h-term estimation. 
	
	To obtain the best-fit value of $A_1$, $A_2$ and $A_3$, we minimize the likelihood 
	\ba
		L \propto \rm exp \left(-\frac 1 2 \chi^2\right),
	\ea
	and 
	\begin{eqnarray*}
		\chi^2 = [\hat y(\vec{\theta}) - y^{\rm th}(\vec{\theta})]^\dagger \textbf{C}^{-1} [\hat y(\vec{\theta}) - y^{\rm th}(\vec{\theta})].
	\end{eqnarray*}
	$\textbf{C}$ is the covariance matrix. We estimate it by Jackknife resampling. The number of the Jackknife sample is set to be $N_{\rm JK} = 100$.
	The Fisher matrix for free parameters $A_1$, $A_2$ and $A_3$ is
	\begin{eqnarray}
		\textbf{F}=
		\left (
		\begin{array}{ccc}
			\textbf{y}^T_1 \textbf{$C^{-1}$}\textbf{y}_1 & 
			\textbf{y}^T_1\textbf{$C^{-1}$}\textbf{y}_2 & 
			\textbf{y}^T_1\textbf{$C^{-1}$}\textbf{y}_3 \\
			
			\textbf{y}^T_2\textbf{$C^{-1}$}y_1 & 
			\textbf{y}^T_2\textbf{$C^{-1}$}\textbf{y}_2 & 
			\textbf{y}^T_2\textbf{$C^{-1}$}\textbf{y}_3 \\
			
			\textbf{y}^T_3\textbf{$C^{-1}$}\textbf{y}_1 & 
			\textbf{y}^T_3\textbf{$C^{-1}$}\textbf{y}_2 & 
			\textbf{y}^T_3\textbf{$C^{-1}$}\textbf{y}^T_3 \\
		\end{array}
		\right),
	\end{eqnarray}
	where 
	\ba
		\textbf{y}_i=\left(  y_i(\theta_1), y_i(\theta_2), ..., y_i(\theta_n) \right)^T
	\ea
	is the template of 1h-, 2h-term or background term. 
	The best-fitting values of $A_1$, $A_2$ and $A_3$ are the solution of equation
	
	\begin{eqnarray}
		\textbf{F} \times
		\left (
		\begin{array}{ccc}
			A_1\\A_2\\A_3
		\end{array}
		\right)
		=
		\left (
		\begin{array}{ccc}
			\textbf{y}_1\textbf{$C^{-1}$}\hat y \\
			\textbf{y}_2\textbf{$C^{-1}$}\hat y \\
			\textbf{y}_3\textbf{$C^{-1}$}\hat y 
		\end{array}
		\right)\,.
	\end{eqnarray}
	And the uncertainty of $A_i$ is
	\ba
		\sigma_i^2 = \textbf (F ^{-1})_{ii}.
	\ea
	The S/N of $A_i$ is defined as $A_i/\sigma_{A_i}$.
	

\section{Analysis}\label{sec:result}
	
	We measure the stacked tSZ profile from \emph{Planck} MILCA map for our 39 DESI cluster/group subsamples. 
	The stacked tSZ profiles as a function of angular radius are shown in Fig.\ref{fig:y_profile}. For brevity, we only show the results of $0.4\leq z <  0.6$. The results of the other three redshift bins are similar. As expected, the stacked tSZ peaks at the center and decreases with  increasing radius. The peak value drops nearly monotonically with decreasing cluster mass. It drops by a factor of 100 from the most massive cluster ($10^{15}M_\odot/h$) to small groups $\sim 10^{13} M_\odot/h$).
	
\begin{figure}
		\centering
 		\includegraphics[width=0.45\textwidth]{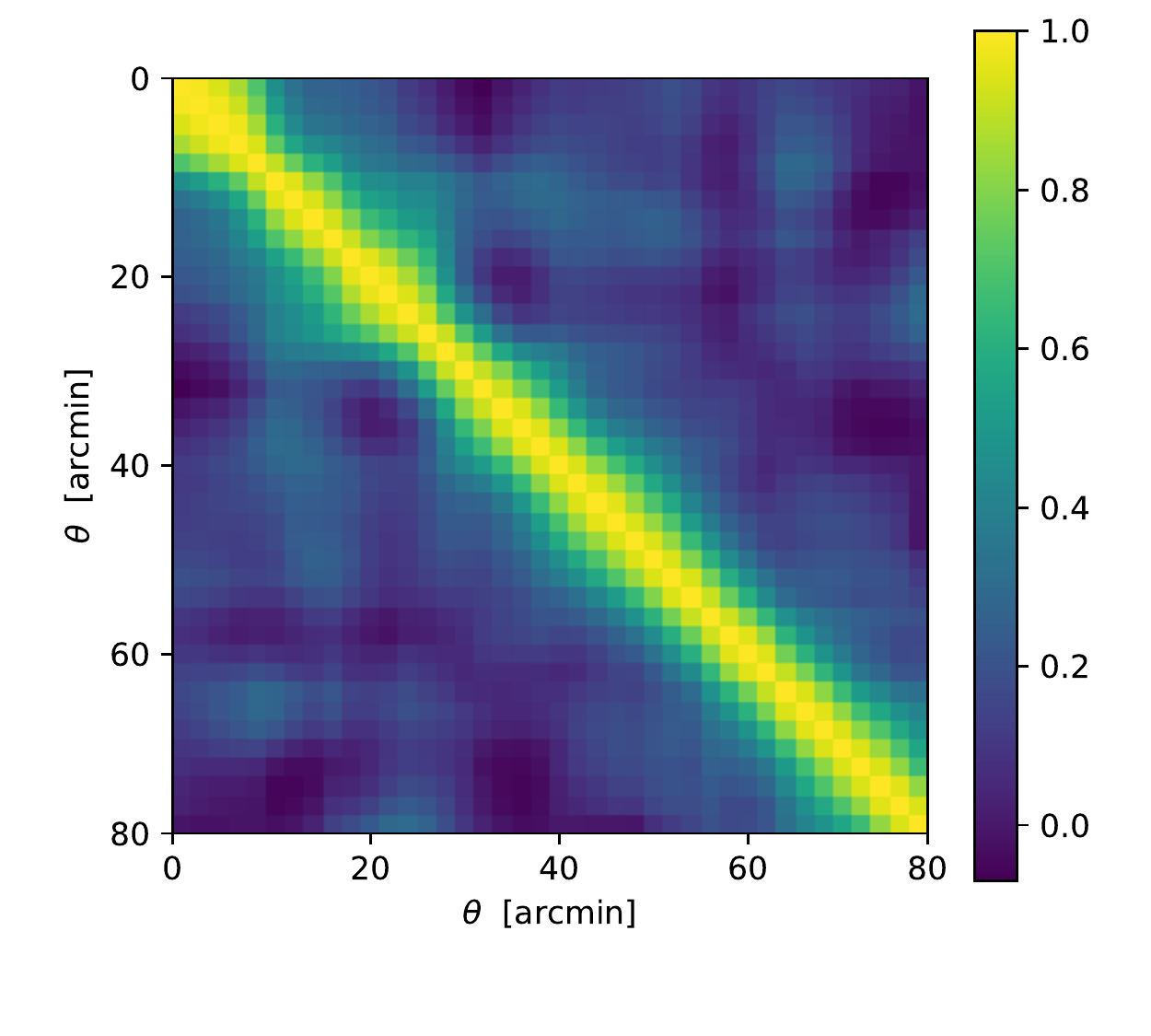}
 		\caption{The normalized covariance matrix of data points shown in Fig.\ref{fig:y_profile}. Here we only show the result of a cluster sample with $0.4 \leq z < 0.6$ and $14.9 < \lg M_L < 15.5$ (top left panel in Fig.\ref{fig:y_profile}). The covariance matrices for other cluster samples are similar. Near data points are strongly correlated, largely due to the \emph{Planck} beam. \label{fig:cov}}
	\end{figure}

	Fig.\ref{fig:cov} shows the normalized covariance matrix for the most massive bin in $0.4\leq z <  0.6$. The covariance of other bins is similar. There is large correlation between neighboring bins separated by $\la 5\ \rm arcmin$. This may be caused by the beam and mis-centering effects which mix the signal at different radius scale. So despite the fact that we have 40 data points for each sample, the really degree of freedom (namely independent data points) is $\sim 10$. For these reason, we choose $\chi^2_{\rm min} \leq 10$ as the criterion of a good fit.

	The combination of a 1h-term template (Eq.\ref{eq:y1}), a 2h-term template (Eq.\ref{eq:y2}) and a constant background term are fitted to the measured tSZ signal. The fitted $A_i$ and their uncertainties are shown in Table.\ref{tab:all}. When the cluster mass is larger than $10^{14.5} M_\odot/h$, the 1h-term dominates the y-profile and the measurement significance of the 1h-term is high.

	\begin{figure}
		\centering
		\includegraphics[width=.45\textwidth]{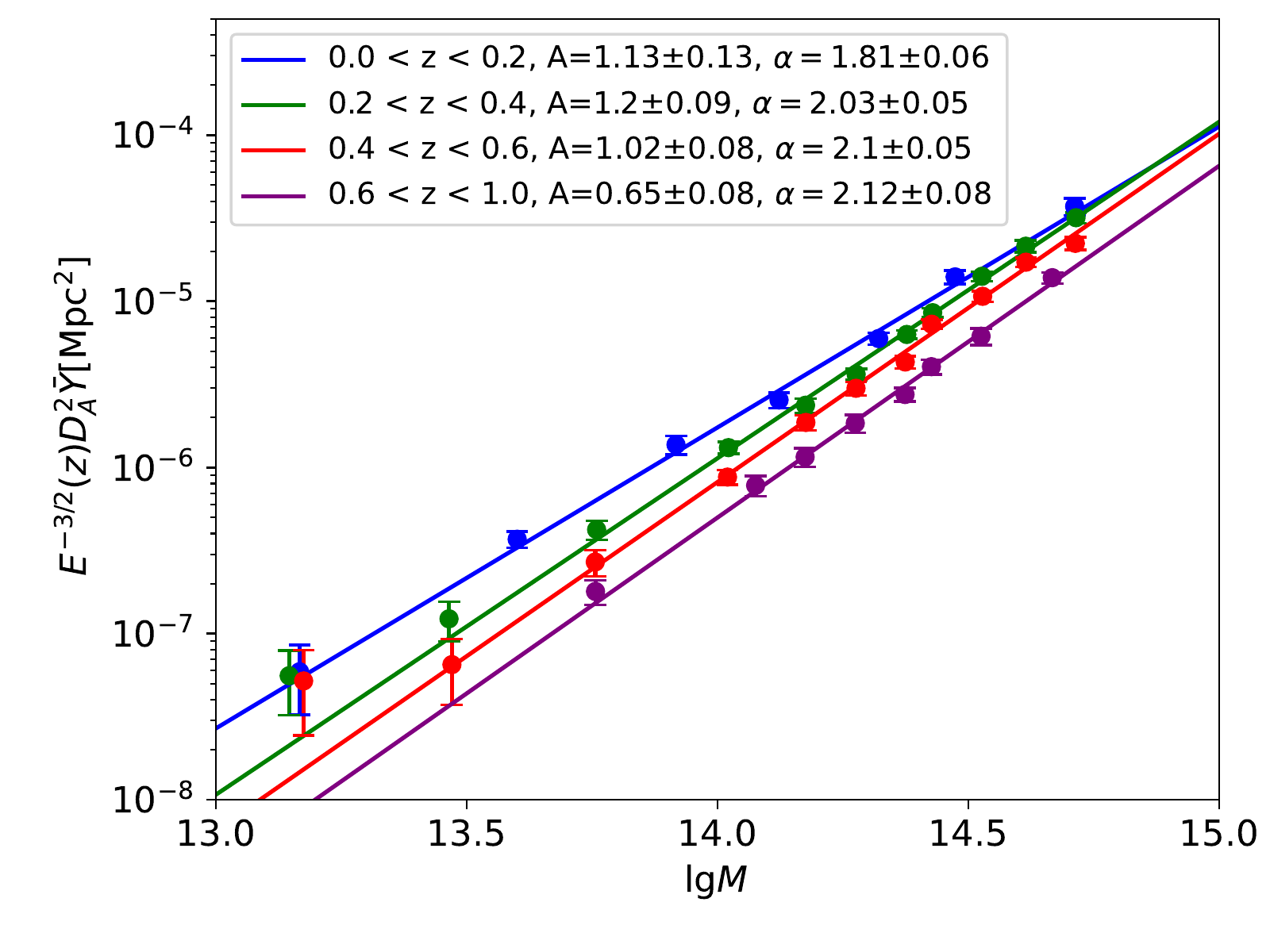}
 		\caption{The relation between parameter $Y$ (Eq.\ref{eq:Y}) and cluster mass $M$. The points with errorbars represent measurement results from 1h-term. These points can described by Eq.\ref{eq:MY} shown as solid lines with parameters $A$ and $\alpha$. Since the data now spans two orders of magnitude in $M$, constraints on alpha are tight. \label{fig:M_Y}}
	\end{figure}

	\subsection{1h-term constraints}
	The 1h-term of tSZ profile is determined by thermal energy distribution in a halo. We utilize its measurement to constrain $Y-M$ relation and find this relation is redshift-dependent. Then we conduct a thermal energy census by integrating all thermal energy in halos of the whole mass range.
	Besides, the coefficient parameter of 1h-term, $A_1$, describes the baryon abundance in clusters. It can capure the decreasing of KS profile at overall r-range. In Appendix. \ref{app:concentration}, we discuss how the change of profile shape would influence the fitting results and find the influence is not obvious ($<1 \sigma$).

	\subsubsection{$Y-M$ relation}\label{subsec:YM}
	
	With the measurement of 1h-term coefficient $A_1$ and template $y_1(\theta)$ (Eq.{\ref{eq:y1}),  we calculate the parameter $Y$\footnote{$y_1(\theta)$ here induces the beam effect, so the integral is $0< \theta < \infty$. This equals to the integral of the original $y_1(\theta)$ to $\theta_{\rm vir}$}
	\ba\label{eq:Y}
		Y=\int A_1y_1(\theta)d^2\theta
	\ea
	as a function of cluster mass in different redshift bins. \add{Note that the amplitude of $Y$ strongly depends on the radius truncation ($R_{vir}$ or $R_{500}$) of the integration.} The $Y-M$ relations of 37 group samples are shown in Fig.\ref{fig:M_Y} as dots with errorbars. The most massive bins of $0.2\leq z <  0.4$, $0.4\leq z <  0.6$ are abandoned, and the reason is explained in Appendix. \ref{app:mtml}.
	
	Over the last decade, quite a lot of SZ-selected samples have been published by \emph{Planck} \citep{ Planck2014_SZ_cosmology, Planck2014_SZ_catalog, Planck2015_SZ_catalog_update}, ACT \citep{Marriage2011, Hasselfield2013}, SPT \citep{Reichardt2013, Bleem2015} and other surveys \citep{Brodwin2015}. Our work probes the $Y-M$ relation down to much lower mass ($10^{13} M_\odot/h \sim 10^{14} M_\odot/h$) than the above samples. In \citep{Planck2014_SZ_cosmology}, the $Y-M$ relation can be described by the formula
	\ba\label{eq:MY}
		E^{-3/2}(z)\left[ \frac {D^2_A Y} {A\times 10^{-4} {\rm Mpc}^2}\right]=\left( \frac{M}{10^{14.5} M_\odot/h} \right)^{\alpha},
	\ea
	where 
	\ba
		E(z)=\sqrt{\Omega_m(1+z)^3+\Omega_\Lambda}
	\ea
	assuming the universe is flat and $D_A$ is the angular distance.
	
	We fit this formula with the measurements and the results are shown as the solid lines in Fig.\ref{fig:M_Y}. Under the hydrostatic equilibrium, the slope $\alpha=5/3$.  
    \add{In our measurements, the slope $\alpha$ exceeds 1.8 in all redshift bins, indicating that less massive clusters possess less pressure than the hydrostatic equilibrium prediction. This can be explained by two possibilities. Firstly, the baryon fraction is lower in less massive halos as supported by our measurements (Sec. \ref{subsec:AR_1h}). Alternatively, it could be caused by a higher non-pressure fraction in these halos, while this trend is not observed in \citet{Shi2015}. Furthermore, we have observed a slight increase in $\alpha$ with the increasing of redshift. However, besides the lowest redshift bin, which is affected by mass determination systematic errors mostly, the redshift evolution is not evident among three other bins. Further measurements are necessary to verify the existence of this redshift evolution.}


	\subsubsection{Thermal energy census of bound gas}
	\begin{figure*}
		\centering
 		\includegraphics[width=\textwidth, height=0.25\textwidth ]{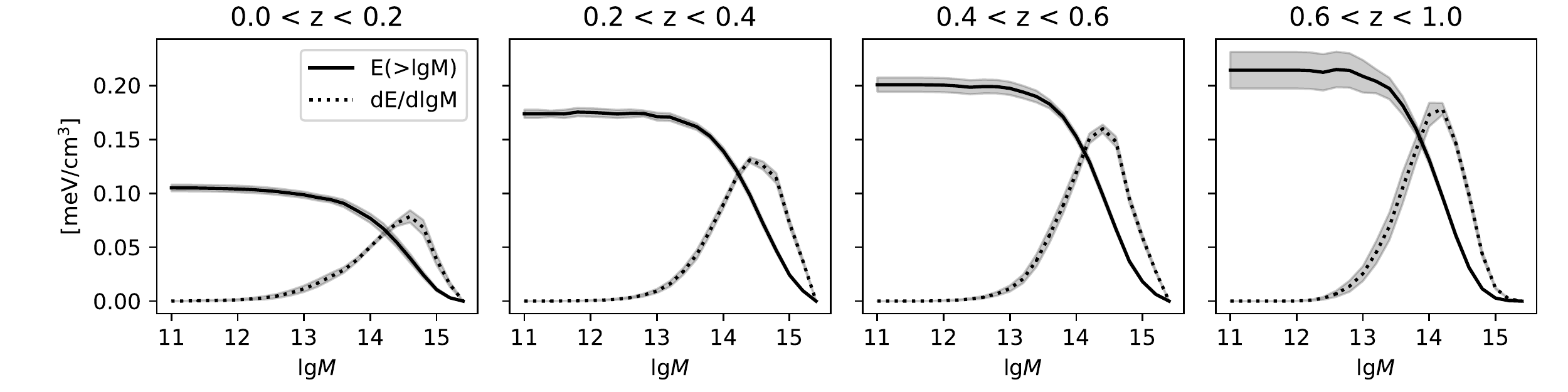}
 		\caption{The thermal energy census of hot baryons bound to clusters/groups. The dashed lines show the thermal energy contributed by halos with different mass. The solid lines show the accumulative thermal energy. The grey shadow region is the uncertainty from the estimation of $\lg M_0$ and $\beta$ in $f_{\rm gas} - \lg M$ relation. \label{fig:thermal_consus}}
	\end{figure*}
	The thermal energy of baryons, although tiny, is an important property of the universe \citep{Zhang2004,Fukugita2004, Chiang2020}.
	The tSZ effect is contributed by all thermal energy in the universe. It is known that most thermal energy resides in the center of massive clusters, which is also observed in IllustrisTNG simulation \citep{Gouin2022}. Thanks to the high completeness of DESI group catalog DR9 which contains almost all massive clusters from z=0 to z=1, a census of thermal energy which bound to halos in the universe can be conducted.
	The relation between thermal gas pressure and the electron pressure measured by the tSZ effect is
	\ba
		P_{\rm th}=\frac{3+5X}{2+2X}P_e,
	\ea
	where X=0.76 is the primordial hydrogen abundance. The profile of $P_e$ is from Eq. \ref{eq:pe} and $f_{\rm gas}(M)$ obtained by 1h-term measurement (Sec.\ref{subsec:AR_1h}). With the assumption that baryons are fully ionized, $P_{\rm th}=1.93P_e$, and the thermal energy 
	\ba
		E_{\rm th}=\frac 3 2 P_{\rm th}.
	\ea
	Therefore the total thermal energy density bound to halos is 
	\ba
		E=\int \left(\int E_{\rm th} dV\right) \frac{dn}{d\lg M}d\lg M.
	\ea
	The inner integration represents the total thermal energy in a halo with mass $M$. $\frac{dn}{d\lg M}$ is obtained from the mass distribution of Y21 catalog.
	
	In Fig.\ref{fig:thermal_consus}, the dashed lines show the contribution of thermal energy from clusters with different masses. Most of the thermal energy is coming from clusters with mass \add{$>10^{13}$.}
	Half of the thermal energy is coming from cluster whose mass is above $10^{14.43}, 10^{14.49}, 10^{14.39}, 10^{14.14} M_{\odot}/h$ in redshift bin $0\leq z <  0.2$, $0.2\leq z <  0.4$, $0.4\leq z <  0.6$, $0.6\leq z \leq 1$. 
	And the peak tends to move to lower mass with the increasing of redshift. 
	The  solid lines in the figure show the accumulated thermal energy as a function of cluster mass. A plateau is reached before $M=10^{12}M_\odot/h$ in all redshift bins, which means the thermal energy in cluster/group with mass smaller than $10^{12} M_\odot/h$ can be ignored. In our measurement, the thermal energy bound to halos is $0.105\pm 0.003$, $0.174\pm 0.004$, $0.201\pm 0.007$ and $0.214\pm 0.017$ [meV/cm$^3$] in redshift bin $0\leq z <  0.2$, $0.2\leq z <  0.4$, $0.4\leq z <  0.6$ and $0.6\leq z \leq 1$.

\subsection{2h-term constraints: $\langle b_yP_e\rangle$} \label{subsec:AR_2h}
	\begin{figure*}
		\centering
 		\includegraphics[width=\textwidth]{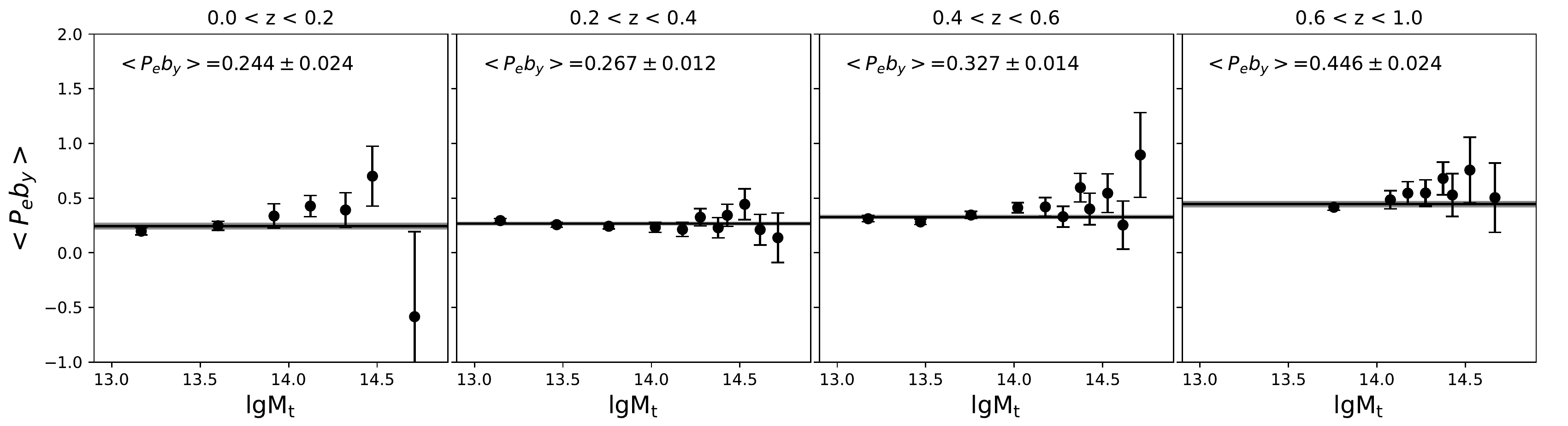}
 		\caption{The bias-weighted electron pressure $\langle b_yP_e\rangle$ as a function of the calibrated cluster/group mass in four redshift bins. The solid line is the mean value estimated from all mass bins in this redshift (Eq.\ref{eq:mean_peby_esitimate}) and the grey shadow region is the corresponding uncertainty. These two are labeled on the top left of each panel. \label{fig:bype}}
	\end{figure*}

	\begin{deluxetable}{cccccc}
	\tablecaption{The best-fitted value of the bias-weighted electron pressure and its uncertainty of each redshift bin.\label{tab:2h}}
	\tablehead{\colhead{range of $z$} & \colhead{$z_{\rm mean}$} &\colhead{$\langle b_yP_e\rangle$ [$m$eV/cm$^3$]} }
	\startdata
    $[$ 0.0, 0.2 )& 0.14 & $ 0.244 \pm 0.025 $\\
    $[$ 0.2, 0.4 )& 0.31 & $ 0.267 \pm 0.013 $\\
    $[$ 0.4, 0.6 )& 0.49 & $ 0.327 \pm 0.015 $\\
    $[$ 0.6, 1.0 )& 0.75 & $ 0.446 \pm 0.024 $\\
    \hline
    $[$ 0.1, 0.2 )& 0.15 & $ 0.200 \pm 0.024 $\\
    $[$ 0.2, 0.3 )& 0.25 & $ 0.222 \pm 0.017 $\\
    $[$ 0.3, 0.4 )& 0.35 & $ 0.301 \pm 0.017 $\\
    $[$ 0.4, 0.5 )& 0.45 & $ 0.302 \pm 0.018 $\\
    $[$ 0.5, 0.6 )& 0.55 & $ 0.395 \pm 0.027 $\\
	\enddata
	\end{deluxetable}

	\begin{figure}
		\centering
 		\includegraphics[width=0.5\textwidth]{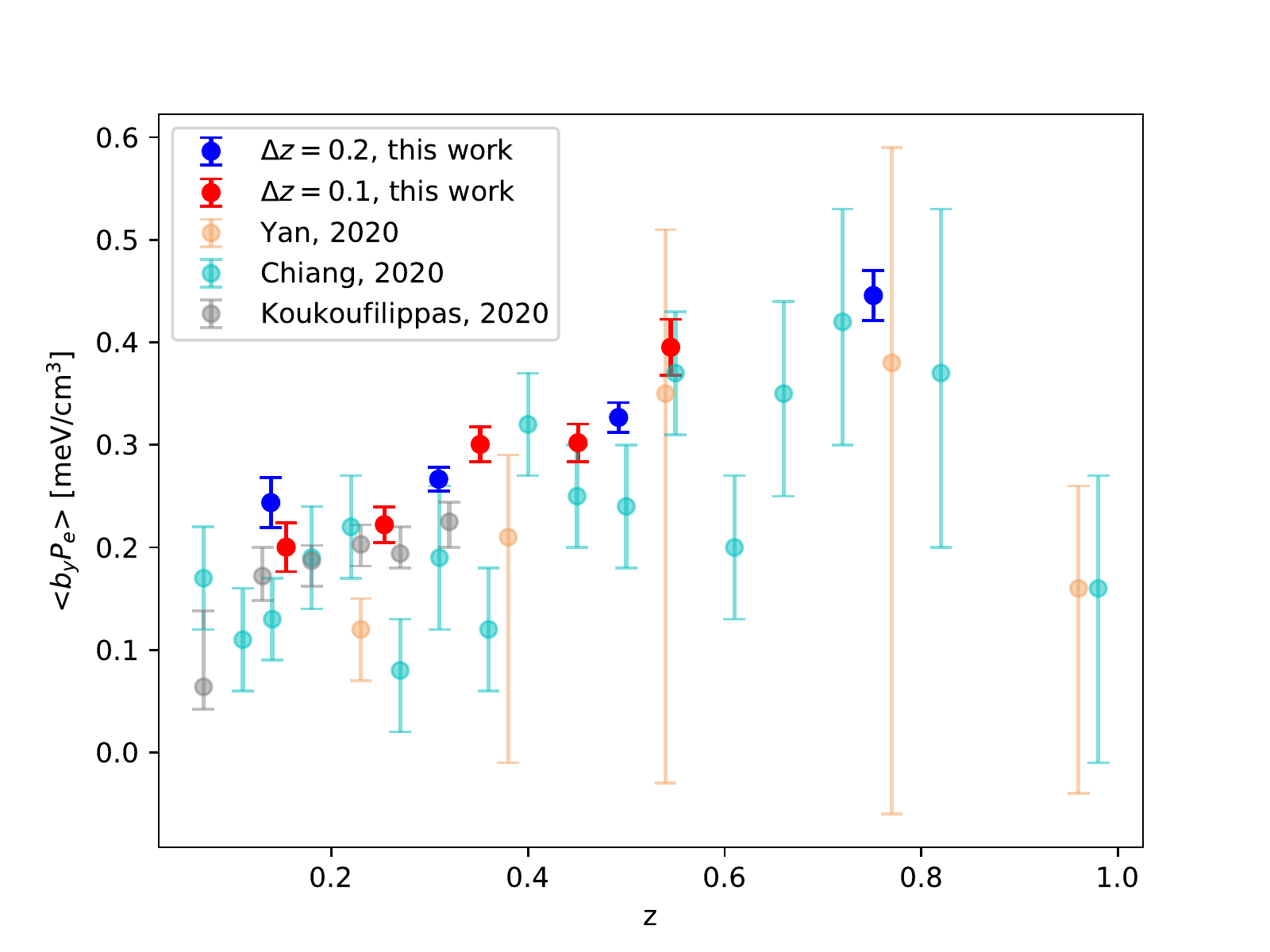}
 		\caption{The comparison of $\langle b_yP_e \rangle$ with other measurements in literature. The blue and red points are our measurements with $\Delta z=0.2$ and $\Delta z=0.1$. The orange, cyan and grey points are the measurements from \citet{Yan2021, Chiang2020, Koukou2020}. All these data points measured by different methods and data sets are consistent. \label{fig:bype_compare}}
	\end{figure}
	
    The coefficient of the 2h-term $A_2$ describes the bias-weighted electron pressure $\langle b_yP_e \rangle$. We show $\langle b_yP_e \rangle$ of different mass and redshift bins in Fig. \ref{fig:bype}. 
    \add{In the fitting we absorb galaxy bias $b_g$ in the 2h-term template and fix its value by a $b_g-M$ relation model.}
    The value of different mass bins in the same redshift bin are consistent with each other in the $1\sigma-$errorbar. 
    \add{This reflects the validity of the $b_g-M$ model}.
    The $\langle b_yP_e \rangle$ estimated from all mass bins is 
	\ba\label{eq:mean_peby_esitimate}
		\langle b_yP_e \rangle=\frac{\sum_i \langle b_yP_e \rangle_i/\sigma_i^2}{\sum_i 1/\sigma_i^2},
	\ea
	where $i$ represents different mass bins in the same redshift bin. Its relationship with redshift is an indicator of cosmic thermal history.
	
	The measured bias-weighted electron pressure as a function of redshift is shown in Fig.\ref{fig:bype_compare}. $\langle b_yP_e\rangle$ increases monotonically with increasing redshift. The redshift interval is $\Delta z=0.2$ in our fiducial measurement. We find that $\langle b_yP_e \rangle$ increases significantly with $z$. Therefore we also try a smaller redshift bin size $\Delta z=0.1$. The uncertainty of each red point is larger than that of each blue point due to the decreasing number of clusters in a narrow redshift bin. Besides, the measurements of different redshift bin lengths are consistent with each other. The measurements of these redshift bins are summarized in Table.\ref{tab:2h}. 
	
	We also show the comparison with other measurements in literature. \citet{Yan2021} adopts the galaxy-tSZ-CMB lensing cross-correlation using \emph{Planck} and \add{Kilo-Degree Survey} data (KiDS; \cite{Kuijken2019}). \citet{Chiang2020} uses the tomographic tSZ measurements from the \emph{Planck} and a spectroscopic galaxy sample from SDSS. \citet{Koukou2020} uses the same method but a photometric redshift galaxy sample from WISE $\times$ SuperCOSMOS public catalog (grey points). In Fig.\ref{fig:bype_compare}, these results from different methods and different data sets are consistent with each other.	
	
\subsection{Consistency of 1h- and 2h-terms}
	\begin{figure}
		\centering
 		\includegraphics[width=0.5\textwidth]{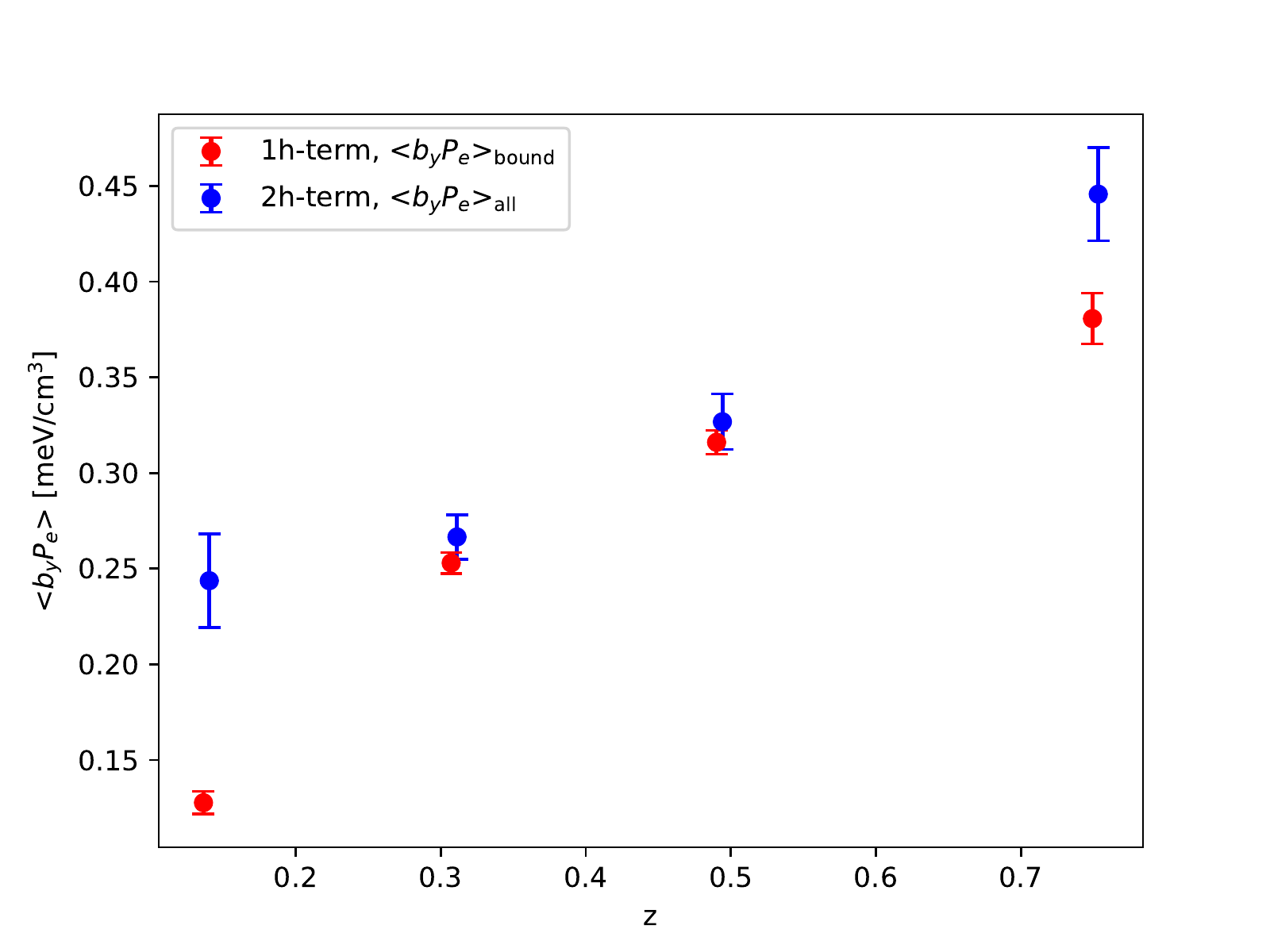}
 		\caption{The comparison of $\langle b_yP_e \rangle$ induced from 1h- and 2h-terms. This plot verifies the consistency relation Eq.\ref{eq:bype_2}. \label{fig:1h2h_consist}}
	\end{figure}

	From the measurement of 1h-term, the electron pressure density profile $P_e(r)$ of a halo with a given mass and redshift can be inferred. TSZ 2h-term (Sec.\ref{subsec:AR_2h}) measures the mean $\langle b_yP_e \rangle$ as a function of redshift. In this subsection, we check the consistency between the 1h- and 2h-term measurements. With the assumption of the halo model, the mean \add{electron} pressure which is bound to halos is 
	\ba
		\langle P_{e}\rangle_{ \rm bound} =\int \left( \int P_e dV \right) \frac{dn}{dM} dM,
	\ea
	and 
	\ba
		b_{y, \rm bound}=\frac{\int b_g \bar Y \frac{dn}{dM} dM}{\int b_g \frac{dn}{dM} dM}.
	\ea
	We adopt the halo mass function and bias-mass relation from \citep{sheth2001}.
	Then we estimated the $\langle b_yP_e \rangle_{\rm bound}$ from 1h-term measurement. The comparison with the measurement from 2h-term is shown in Fig.\ref{fig:1h2h_consist}. $\langle b_yP_e \rangle$ measured by the two halo term is the sum of that of bound and unbound gas
	\ba
		\langle b_yP_e \rangle = \langle b_yP_e \rangle_{\rm bound} + \langle b_yP_e \rangle_{\rm unbound}.
	\ea
	 Therefore, 
	 \ba\label{eq:bype_2}
	 \langle b_yP_e \rangle >\langle b_yP_e \rangle_{\rm bound}\ .
	 \ea
	 Eq.\ref{eq:bype_2} is the consistency relation that we need to test  with the both 1h- and 2h-term measurements.
	 Fig.\ref{fig:1h2h_consist} shows that this relation is indeed satisfied at all four redshift bins. We will further discuss the difference between $\label{bype_2}
	 \langle b_yP_e \rangle$ and $\langle b_yP_e \rangle_{\rm bound}$ in Sec. \ref{subsec:ejectedP}.
	 
\section{astrophysical and Cosmological implications}\label{sec:appli}

	\subsection{Baryon abundance in clusters}\label{subsec:AR_1h}
	
	\begin{figure*}
		\centering
 		\includegraphics[width=\textwidth]{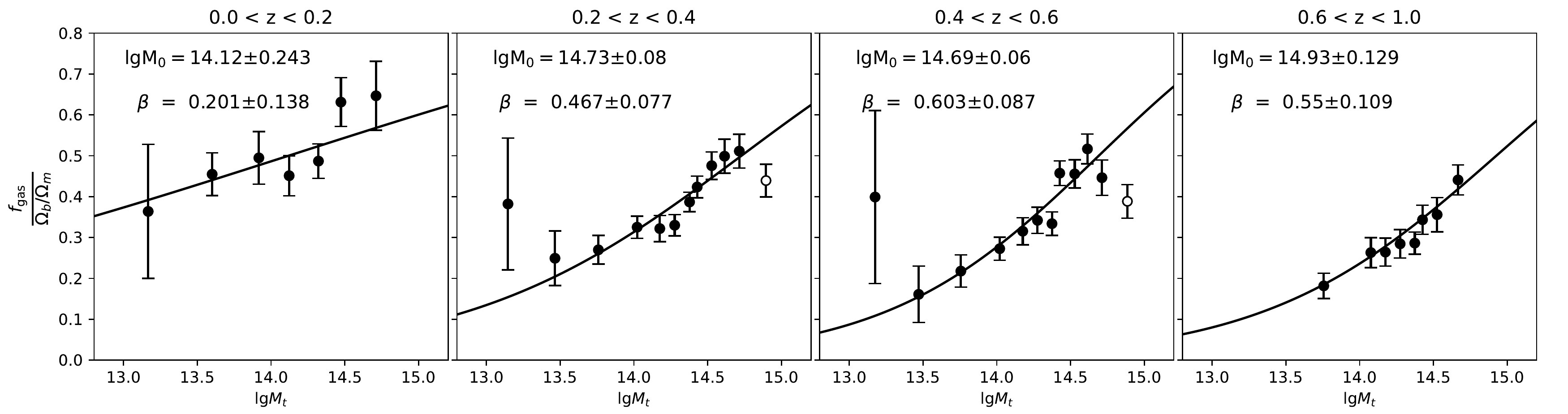}
 		\caption{The gas abundance as a function of cluster mass in four redshift bins. In each panel, the black points are the fitting coefficients of the 1h-term template. The solid line is the fitting of Eq.\ref{eq:fgas} with these black points, while the hollow circle points are not included in the fitting. The best-fitting values and uncertainties of the parameters $\lg M_0$ and $\beta$ are shown on the top-left corner of the panel. \label{fig:fgas}}
	\end{figure*}
	
	 The baryon feedback, such as SN and AGN, would heat/blow the gas out of the halo. The potential suppression of gas fraction is captured by the parameter $A_1$\footnote{This interpretation of $A_1$ relies on the condition that the modeling of $y_1(\theta)$ is accurate. To be specific, the validity of the $A_1$ interpretation assumes that feedback would not change the temperature profile. This assumption is valid at the first order in which the temperature is fixed by the gravitational potential. But at high order, this interpretation may be inaccurate.}. Given $A_1$ measurement in Table.\ref{tab:all}, we obtain the relationship between $A_1$ and the cluster mass $M_t$ at each redshift.  We fit this relation against the formula \citep{Schneider2015}
	\ba\label{eq:fgas}
		A_1=\frac{f_{\rm gas}}{\Omega_b/\Omega_m}=\frac{(M/M_0)^\beta}{1+(M/M_0)^\beta}\ , 
	\ea
	where $\lg M_0$ and $\beta$ are the  parameters to fit. In this parameterization, the gas fraction of a cluster with $M=M_0$ is $50\%$ of the cosmic mean. 
	
	The fitting results are shown in Fig.\ref{fig:fgas}. We discard the the most massive bin of $0.2\leq z <  0.4$ and $0.4\leq z <  0.6$, because their mass distribution may be modeled inaccurately (Appendix.\ref{app:mtml}). The best-fitting values and the uncertainties of $\lg M_0$ and $\beta$ are also shown in Fig.\ref{fig:fgas}.
	
	The baryon abundance becomes lower as the decreasing of the halo mass in each redshift bin. 
	$M_0=10^{14.12, 14.73, 14.69, 14.93} M_\odot/h$  at redshift bin $0\leq z <  0.2$, $0.2\leq z <  0.4$, $0.4\leq z <  0.6$ and $0.6\leq z \leq 1$.
	Correspondingly, $\beta = 0.20\pm 0.14, 0.47\pm 0.08, 0.60\pm 0.09, 0.55\pm 0.11$. 
	These results imply a strong impact of feedback on cluster gas fraction. In addition, we find a slight evolution of cluster baryon abundance with redshift. The baryon abundance is higher in low redshift than high redshift in halos with the same mass. We will verify these findings in a future work, especially with the help of hydronamical simulations.

	In previous works, halo baryon abundance is usually measured by X-ray observation.
	\citet{Sun2009} finds a average value $f_{\rm gas}\sim 0.12$ and the slope of $f_{\rm gas}-M_{500}$ relation is $0.135\pm 0.030$ with large scatter, using 43 nearby galaxy groups with $10^{13} < M_{500} < 10^{14} M_\odot/h$, $0.012 < z< 0.12$ based on \emph{Chandra} archival data.
	With 49 low-redshift clusters provided by \emph{Chandra} and \add{ROSAT} data, \citet{Vikhlinin2009} fits a linear relation of $f_{\rm gas}$ and $\lg M_{500}$ with the mass range $M_{500} > 13.7 M_\odot /h$. 
	In addition, \citet{Gonzalez2013} finds a $f_{\rm gas}-M_{500}$ relation $f_{\rm gas} \propto M_{500}^{0.26\pm 0.03}$  for 12 galaxy groups/clusters at $z\sim 0.1$ with $10^{14} < M_{500} < 5\times 10^{14} M_\odot/h$ observed by \emph{XMM-Newton} X-ray telescope. There is distinguishable difference in slope for cluster samples with different mass range.
	Utilizing all above X-ray measurements, \citet{Schneider2015} parametrise the $f_{\rm gas}-M_{500}$ relation as Eq. \ref{eq:fgas} with the best-fitting parameters $M_0=1.2\times 10^{14} M_\odot /h$ and $\beta = 0.6$.
	$\beta =0.6$ is different from our measurement $\beta =0.20\pm 0.14$ in $0 \leq z < 0.2$. This may be caused by the poor description of the measurement by Eq. \ref{eq:fgas} (left panel in Fig. \ref{fig:fgas}).
	Considering the large measurement uncertainty and different mass definition, however, these X-ray measurements are consistent with our results in the lowest redshift bin. 
	Also, \citet{Lim2018} constrains the hot gas fraction using tSZ effect with the combination of \emph{Planck} and a group catalog given in \citet{Lim2017} at redshift $z<0.2$. Although they only adopt low redshift cluster samples, in the halo mass range $M_{200} > 10^{13.5} M_\odot /h$ they find a similar relation with ours. 
	
	Recently \citet{chen2022} constrained the baryonic feedback combining The DES Year-3 small scale cosmic shear measurement and the baryon correction model (BCM), in which the gas fraction has the same parametrization as Eq. \ref{eq:fgas} \citep{chen2022, Schneider2015, Arico2020}.   The work adopted $\beta=0.321$, which agrees with our constraint at low redshift. It constrained $\log  M_0=14.12^{+0.62}_{-0.37}$, which also agrees with our constraint. 
	
    In addition, we caution that the above obtained gas fraction in dark matter halos, $f_{\rm gas}$, depends on our model assumption of the hydrostatic equilibrium of gas temperature (e.g., Eq. \ref{eq:Tg}), and in general it can be regarded as the `hot' gas fraction. In case that the gas temperature is much lower than the virial temperature of the dark matter halos, the $f_{\rm gas}$ will be underestimated. Very interestingly, in a recent paper, using the kSZ effect around 40000 low redshift groups in the SDSS observation, \citet{Lim2020}  claimed the detection of the ``missing baryons". The total kSZE flux within halos estimated implies that the gas fraction in halos is about the universal baryon fraction, even in low-mass halos with mass $\sim 10^{12.5}M_\odot /h$. It thus indicates that the gas temperature is indeed significantly lower than the halo virial temperature.

	\subsection{Pressure from unbound gas} \label{subsec:ejectedP}
		
	When the gas is ejected from the halo, its fate is  hard to measure and hard to model. The feedback process injects thermal or kinematic energy into these gas. They may lose the energy when escaping from the gravitational wall of the halo or still stay hot. Some models treat these gas just as a diffuse background and only contribute to the 2h-term. And some other models think these gas would not be driven too far away from the halo and reside around the halo, as the so-called \add{Circumgalactic} Medium (CGM) \citep{Tumlinson2017}. Due to the low density of these unbound gas, the measurement of them is very difficult. On the other hand, detecting such gas would learn valuable information on the feedback process. 
	
	Since we have simultaneously measured both the total electron pressure $\langle b_yP_e \rangle $ from the 2-halo term, and  $\langle b_yP_e \rangle _{\rm bound}$ of gas bound to halos from the one halo term, we can directly infer the electron pressure contributed by unbound gas.
	\ba\label{eq:bype_eject1}
		\langle b_yP_e \rangle _{\rm unbound} = \langle b_yP_e \rangle^{2h} - \langle b_yP_e \rangle ^{1h}_{\rm bound}.
	\ea

	\begin{deluxetable}{cccccc}
	\tablecaption{The $\langle b_yP_e\rangle$ contributed by unbound gas and the prediction from scenario 1 \& 2. \label{tab:eject}}
	\tablehead{\colhead{range of $z$} &\colhead{$\langle b_yP_e\rangle_{\rm unbound}$ [$m$eV/cm$^3$]} &\colhead{scenario 1} &\colhead{scenario 2}}
	\startdata
	$[$ 0.1, 0.2 )& $ 0.065 \pm 0.025$ &  0.162 &  0.046 \\
    $[$ 0.2, 0.4 )& $ 0.014 \pm 0.013$ &  0.491 &  0.118 \\
    $[$ 0.4, 0.6 )& $ 0.011 \pm 0.016$ &  0.835 &  0.183 \\
    $[$ 0.6, 1.0 )& $ 0.065 \pm 0.028$ &  1.693 &  0.336 \\
	\enddata
	\end{deluxetable}

The $\langle b_yP_e \rangle _{\rm unbound}$ in different redshift bins are shown in Table.\ref{tab:eject}. We fail to detect $\langle b_gP_e\rangle_{\rm unbound}$ with high significance. However, at least for $z\sim 0.1-0.2$ and $z\sim 0.6-1.0$ we have preliminary detection at $2-3\sigma$, which should be verified with further investigations. Nevertheless, this exercise points out the stacking tSZ measurement is a promising probe of the thermal energy by these diffuse gas out of the halo. For these reasons we present two scenarios and compare them with the measurement. In scenario 1, We assume the unbound gas would accumulate around the halo and do not lose thermal energy when  escaping the halo. So the $b_y$ and $P_e$ of these unbound gas are roughly identical to those bound to halos. Then,
	\ba
		\langle b_yP_e \rangle_{\rm unbound}^{\rm s1}=\int\frac{\Omega_b/\Omega_m-f_{\rm gas}}{f_{\rm gas}}\frac{d\langle b_yP_e \rangle_{\rm bound}}{d\lg M} d\lg M.
	\ea 
	In scenario 2, we assume that the unbound gas keeps its thermal energy, but diffuse into a smooth background. Therefore $b_y=1$. In this case,
	\ba
		\langle b_yP_e \rangle_{\rm unbound}^{\rm s2}=\int\frac{\Omega_b/\Omega_m-f_{\rm gas}}{f_{\rm gas}}\frac{d\langle P_e \rangle_{\rm bound}}{d\lg M} d\lg M.
	\ea
	$\langle b_yP_e \rangle$ of these scenarios are summarized in the third and forth columns of Table.\ref{tab:eject}. For $z>0.2$, predictions of both scenarios are larger than the measurement by a factor of $\ga 10$. Since in scenario we have set $b_y=1$ for unbond/ejected gas, it can be inferred that the temperature of the ejected gas would decrease largely when they escape the halo.\footnote{ The expectations are the lows redshift bins of $0\leq z <  0.2$ and $0.1  \leq z <  0.2$. Since there are other potential problems of group identification and determination of group mass and redshifts, we postpone further investigation on their $\langle b_yP_e\rangle$ until the above problems are significantly improved. }

	\subsection{The suppression of weak lensing power spectrum}

	In this subsection, we show another cosmological implication of tSZ measurement. The probe, weak lensing, is sensitive to the matter distribution and can constrain the parameters of cosmological models.
	The underlying matter power spectrum of a cosmology model is usually provided by dark-matter-only simulations with assuming baryon processes do not impact large-scale structure formation. However, this assumption is no longer valid when $k$ larger than $\sim 1 h/ \rm Mpc$. And the next generation surveys would provide an one-percent level constraint on weak lensing measurements on these scales. Therefore, it is important to characterize the influence on matter power spectrum caused by baryon processing.\\
	Nowadays, a large set of hydrodynamic simulations are used to characterize the suppression of matter power spectrum caused by baryon and its feedback effects \citep{Semboloni2011, Chisari2018, Schneider2019, Debackere2020}. 
	\citet{Harnois2015} constructs an analytic fitting formula that describes the effect of the baryons on the mass power spectrum based on three scenarios of the OWL simulations. 
	\citet{Giri2021} finds the suppression reaches a maximum of 20-28 percent at around $k\sim 7 h/\rm Mpc$ and produces an emulator of baryon effects on the matter power spectrum. 
	\citet{vanDaalen2020} utilizes a set of 92 matter power spectra from several hydrodynamic simulations to conduct a detailed investigation of the dependence on different $\Lambda$CDM cosmologies, neutrino masses, sub-grid prescriptions, and AGN feedback strengths. And they find effectiveness of AGN feedback significantly influence the matter power spectrum on the scale scales $ k > 0.1 h/ \rm Mpc$.
	The results of different hydrodynamic simulations are is determined by the parameters of baryon feedback processes, such as the strength of the AGN feedback. In this subsection, we provide a constraint on how baryon effect would suppress the matter power spectrum from tSZ observation.
	
	\begin{figure}
		\centering
 		\includegraphics[width=0.5\textwidth]{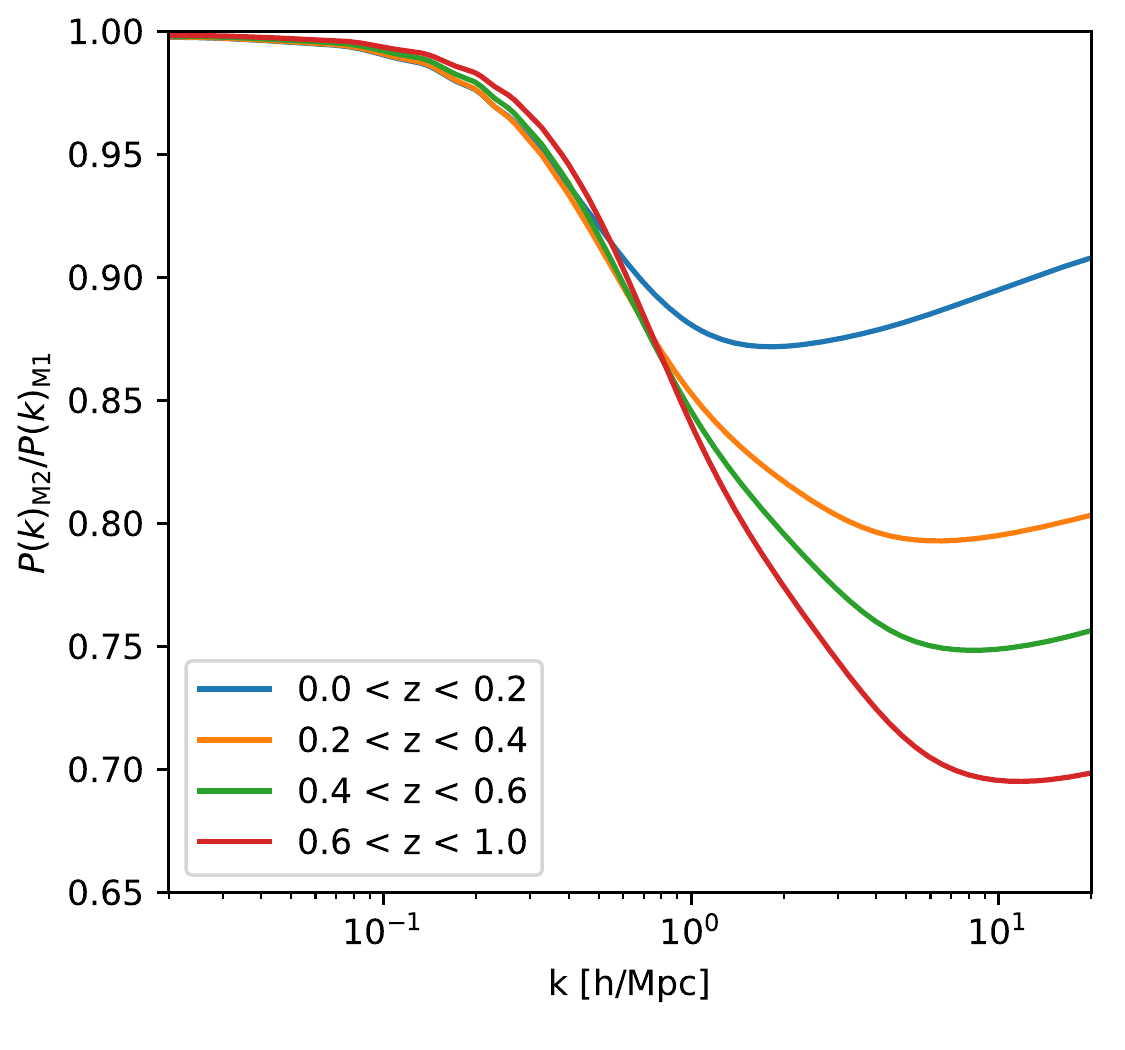}
 		\caption{The estimated suppression of matter power spectrum due to the loss of baryons in clusters/groups. The solid lines are the ratio of matter power spectrum of model 2 (with baryon feedback) and model 1 (without baryon feedback) in four redshift bins. \label{fig:r_ps}}
	\end{figure}
	\begin{figure}
		\centering
 		\includegraphics[width=0.5\textwidth]{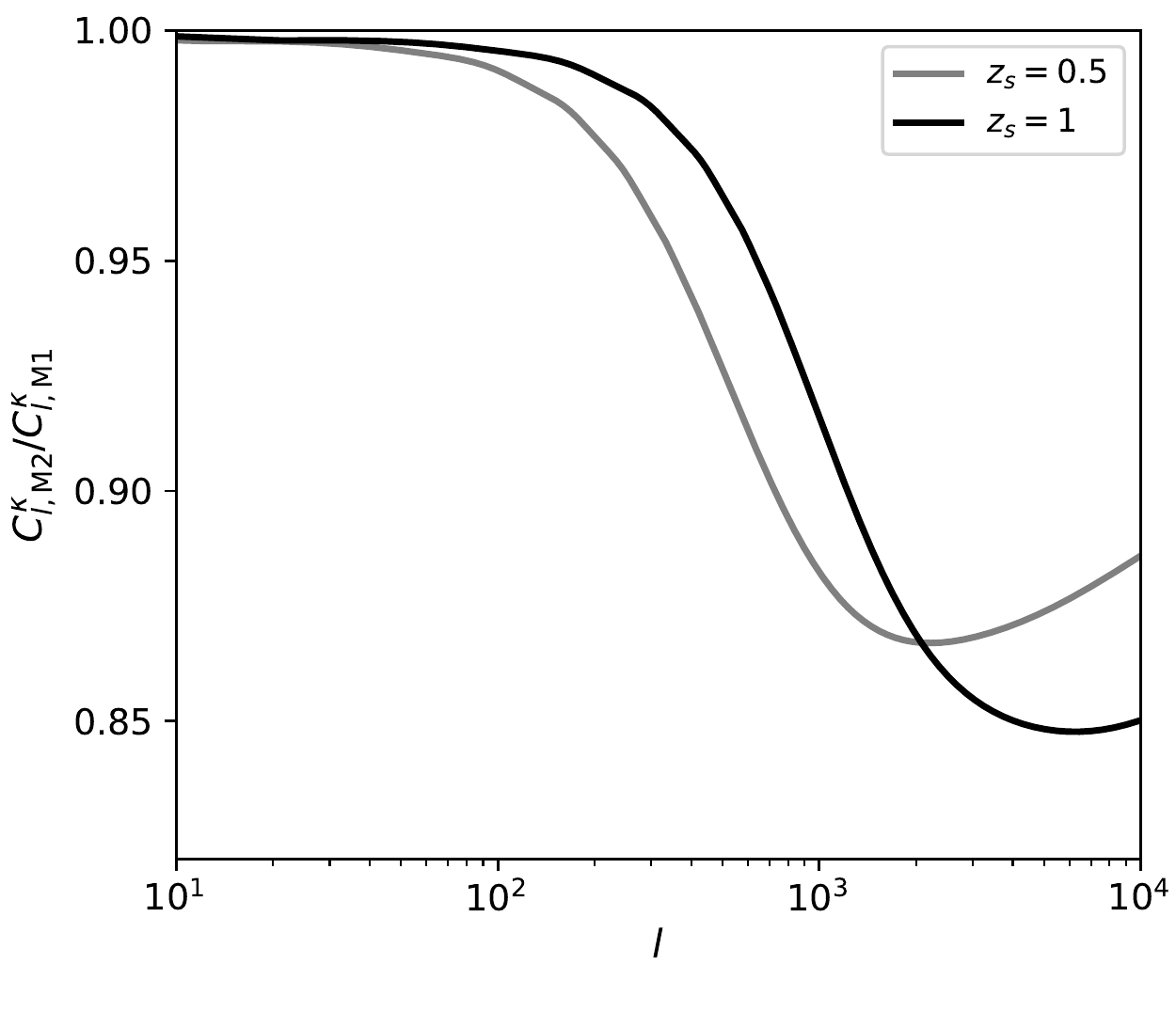}
 		\caption{The estimated suppression of the weak lensing power spectrum due to loss of baryons in clusters/groups.  The lensed source is at $z_s=0.5$ (grey) and $z_s=1$ (black). \label{fig:r_wl}}
	\end{figure}
	
	We utilize the measurement in Sec.\ref{subsec:AR_1h} and adopt the halo model from \citet{Mead2020}. Here is a brief description of the halo model to estimate the lensing statistics. The total power spectrum is the sum of a 1h- and 2h-term 
	\ba
		P_{2h, \mu\nu}(k)&=&P_{\rm lin} \prod \limits_{i=\mu,\nu}[\int b(M)W_i(M,z)n(M)dM] \\ 
		P_{1h, \mu\nu}(k)&=& \int W_\mu(M,z)W_\nu(M,z)n(M)dM,
	\ea
	where $P_{\rm lin}$ is the linear matter power spectrum, $b(M)$ is the linear halo bias, $n(M)$ is the halo mass function, and $\mu, \nu$ represent different matter components such as dark matter, bound gas, and unbound gas.
	For dark matter, we adopt the 'NFW' profile 
	\ba
		\rho_{\rm DM}\propto \frac{1}{r/r_s(1+r/r_s)^2}
	\ea 
	\citep{Navarro1997}. For gas, we adopt a KS profile
	\citep{Komatsu2001}.
	The normalization of these profiles is determined by 
	\ba
		f_i(M)M=\int^{r_v}_0 4\pi r^2\rho_i(M,r)dr.
	\ea
	$f_{\rm dm}$ equals to $({\Omega_{m}-\Omega_{ b}})/{\Omega_m}$. For gas, we set two models to characterize the impact of baryon feedback. For model 1, $f_{\rm bound}={\Omega_{b}}/{\Omega_{m}}$ and $f_{\rm unbound }=0$ which is without baryon feedback. And for model 2, $f_{\rm bound}=f_{\rm gas} $ measured in Sec.\ref{subsec:AR_1h} as a function of mass and redshift and $f_{\rm unbound}={\Omega_{b}}/{\Omega_{m}} - f_{\rm gas}$. We adopt the approximation that the unbound gas does not contribute to the 1h-term, but only to 2h-term as a diffused background.
	
	We show the ratio of matter power spectrum of model 2 and model 1 in Fig.\ref{fig:r_ps}. This ratio quantifies the impact of feedback. At $k=1 h/{\rm Mpc}$, the matter spectrum is suppressed 10\% at z=0 and increasing to 30\% at z=1. The suppression at smaller scales is larger. 
	
	We also show the suppression on weak lensing angular power spectrum, when the source is at $z_s=0.5$ or $z_s=1$ in Fig.\ref{fig:r_wl}. The suppression is about $\ga 10\%$ when $l \ga 1000$. This suppression is an order of magnitude larger than the precision of the weak lensing measurement by Stage IV. This indicates the baryon feedback effect needs to be taken into account in weak lensing measurement. Otherwise, it would become a serious systematic effect entering cosmology constraint from weak lensing measurement with the stage IV survey, such as EUCLID \citep{Laureijs2011}, LSST \citep{LSST2009} and WFIRST \citep{Spergel2015}. Current cosmic shear analysis often mitigates the baryonic effect with scale cuts or adopt models of baryonic effects (e.g \citet{chen2022} on DES year-3 cosmic shear). Independent constraints of baryonic effect from the tSZ measurement will then be highly complementary to correct this effect for weak lensing surveys. 
	
\section{Discussion and conclusion}\label{sec:conclu}
	
	In this work, we utilize the \citet{Yang2021} cluster sample from DESI group catalog DR9 and \emph{Planck} MILCA y-map to measure the tSZ signal. With 0.8 million galaxy clusters/groups and reasonable mass estimation and completeness, we are able to measure both 1h-term and 2h-term with high S/N. The 1h-term measurement provides a differential description of the cluster/group thermal energy, while the 2h-term provides an integral constraint on the thermal energy of all hot baryons, bound and unbound. 
	The 1h-term measurements extend the $Y-M$ relation by one order of magnitude in the mass range. We further find the sign of departure in the $Y-M$ redshift evolution from the prediction of adiabatic gastrophysics. 
	The 2h-term measurements are consistent with previous works, but with smaller errorbars. The comparison between 1h- and 2h-terms provides clue to unbound gas and the impact of feedback. 
	An important cosmological implication is the significant suppression of the weak lensing auto power spectrum which is $\gtrsim 10\%$ at $l\gtrsim 1000$. This confirms the baryonic effect as a major systematic effect in weak lensing.
	
	Although the total S/N of the tSZ detection exceeds 70, our measurement and theoretical interpretation suffer from a number of uncertainties. It is beyond the scope of this work to fully account for these uncertainties in the analysis, due to complexities in describing them and incapability of constraining them by the data. Instead, we list the major uncertainties and discuss the improvements that will be achieved by upcoming surveys. 
\begin{itemize}

\item \textbf{Redshift uncertainty.} The redshift of galaxies in DESI group catalog DR9 is photometric and its uncertainty is about 0.01(1+z) \citep{Yang2021}. The uncertainty of redshift would cause redshift uncertainties of clusters and result in biased templates of 1h- and 2h-term, especially in low redshift bins. The ongoing DESI spectroscopic survey will directly provide spectroscopic redshifts for a fraction of member galaxies. For the rest, cross-correlation between the groups and spectroscopic galaxies will tightly constrain mean redshift of group samples and possibly the outlier rate. 
	
\item \textbf{Mass uncertainty.} The mass uncertainty for DESI group catalog DR9 is 0.2 dex at the high-mass-end and 0.40 dex at the low-mass-end.  In Appendix.\ref{app:mtml}, we use mock data from the simulation to calibrate the mass of clusters in the catalog. However, the number of halos in the simulation is limited, which would cause uncertainty on the $\lg M_t-\lg M_L$ relation. Furthermore, the halo mass function in the simulation relies on the reference cosmology. This may induce certain model dependence in the measurements. The mass uncertainty can be calibrated against cross-correlation with cosmic shear (e.g the catalog\footnote{\url{https://gax.sjtu.edu.cn/data/DESI.html}} constructed by the Fourier-Quad method \citep{Zhang2008}) or spectroscopic galaxies.

\item 	\textbf{The halo concentration.} In Sec.\ref{sec:method}, we assume the concentration of halo is the same as the dark-matter-only situation. However, if a large fraction of baryon is blown away from a halo, the halo would become less compact, corresponding to a smaller concentration. Then the 1h-term profile would be changed. In Appendix.\ref{app:concentration}, we test how the change of concentration would influence our results. Fortunately, we find the influence can't be distinguished within the errorbar, largely due to the poor \emph{Planck} angular resolution. On the other hand, it means the current data are not accurate enough to constrain the halo concentration. Stacking cosmic shear around these groups will constrain not only the total mass but also the concentration-mass relation \citep{wang2022}.

\item 
	\textbf{Non-thermal pressure and baryon feedback} There are two effects would cause difference between the KS profile and the true one. In the first, non-thermal motion, referring as 'turbulence', inside clusters would provides extra pressure support against gravity \citep{shaw2010,shi2014,Osato2018}. The non-thermal fraction $f_{\rm nth}=P_{\rm nth}/(P_{\rm th}+P_{\rm nth})$ monotonously increases with cluster radius and reaches $\sim 0.4$ when $r=r_{200}$ \citep{Nelson2014, Shi2015, Shi2016}. In addition, the baryon feedback would break the hydrodynamic equilibrium and cause departure from KS profile for a fraction of clusters. \add{The thermal pressure profile has also been measured in some previous works \citep{Arnaud2010, Tramonte2023} utilizing generalized NFW (gNFW) formula. However, they are usually using halo mass definition $M_{500}$. Transforming the halo mass definition from $M_{500}$ to $M_{\rm vir}$ is non-trivial, as the baryon processes would alter the NFW mass profile of clusters. Then we will adopt these gNGW profiles in future analysis with a more meticulous calibration.} In this work, we set a free parameter $A_1$ to capture the change of profile amplitude. For the shape, the beam size of Planck is so large that the details of profile shape are smoothed greatly. As checked in Appendix. \ref{app:concentration}, the change of template would not cause a distinguishable difference of the results. \add{}
	
\item 
	\textbf{Mis-centering.} To account for mis-centering of clusters, we set $\eta_{\rm mc}=0.2$ in Eq.\ref{eq:mc} for our fiducial measurement. In appendix.\ref{app:miscentering}, we discuss how the parameter $\eta_{\rm mc}$ would influence the results and find $\eta_{\rm mc}=0.2$ is an optimal choice. Further analysis may adopt more complicated and more realistic description of mis-centering (e.g \citet{Yan2020}). With higher resolution CMB experiments such as ACT, SPT, and CMB-S4, the tSZ data alone will have constraining power for both $c$ and the mis-centering effect.
	
\item 	
	\textbf{Residual foregrounds in tSZ map.} This work adopts \emph{Planck} y-map. In the future, we may follow \citet{Chiang2020} to include non-\emph{Planck} measurements in infrared bands and construct better cleaned y-maps.
\end{itemize}
	
\section{acknowledgments}

This work is supported by the national science foundation of China (Nos. 11621303, 11833005, 11890692), National key R\&D Program of China (Grant No. 2020YFC2201602), CSST CMS-CSST-2021-A02, 111 project No. B20019, and Shanghai Natural Science Foundation, grant Nos. 15ZR1446700 and 19ZR1466800.

This work made use of the Gravity Supercomputer at the
Department of Astronomy, Shanghai Jiao Tong University.

Softwares: astropy \citep{Astropy2013}, numpy \citep{Harris2020}, matplotlib \citep{Hunter2007}, scipy \citep{Virtanen2020}, hmf \citep{Murray2014}, healpy \citep{Zonca2019}

\appendix

\section{The relationship between the true mass and the observed mass}\label{app:mtml}

	\begin{figure*}
		\centering
 		\includegraphics[width=\textwidth]{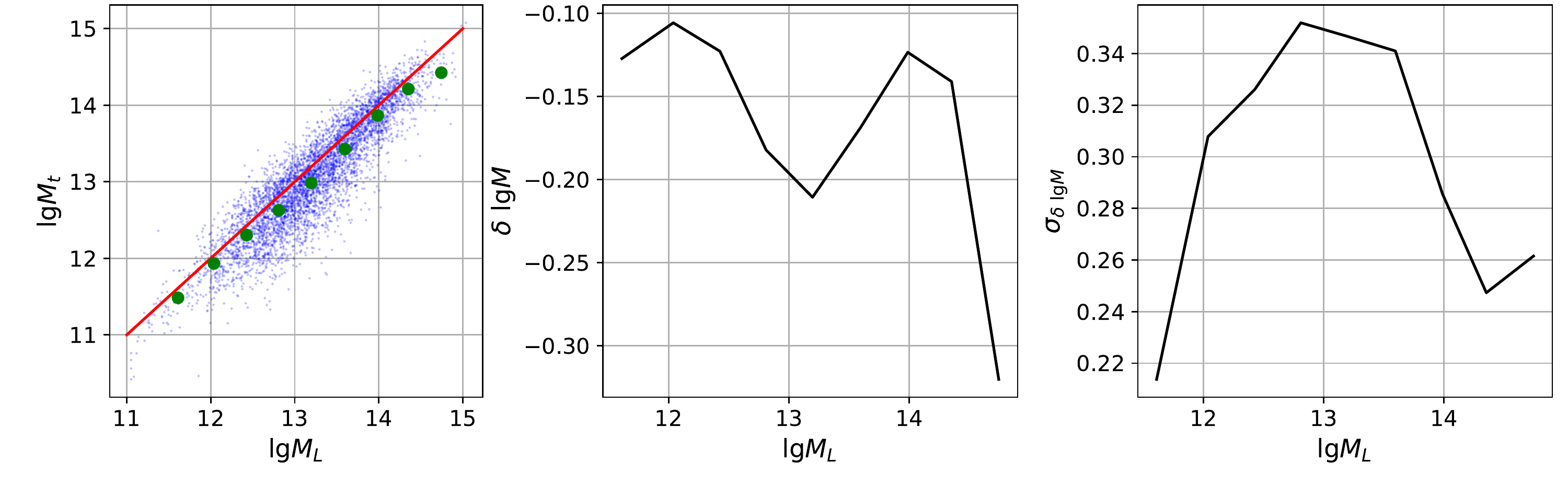}
 		\caption{ \textbf{Left}: each blue point represents a halo in the Mock catalog, the x- and y-axis is the assigned mass and the true mass\label{fig:mlmt}. Red line is $y=x$ and green is the mean $\lg M_t$ as a function of $\lg M_L$. \textbf{Middle}: The mean value of the difference $\delta \lg M$ as a function of $\lg M_L$. \textbf{Right}: The scatter of the difference $\delta \lg M$ as a function of $\lg M_L$. }
	\end{figure*}

	\begin{figure*}
		\centering
 		\includegraphics[width=\textwidth]{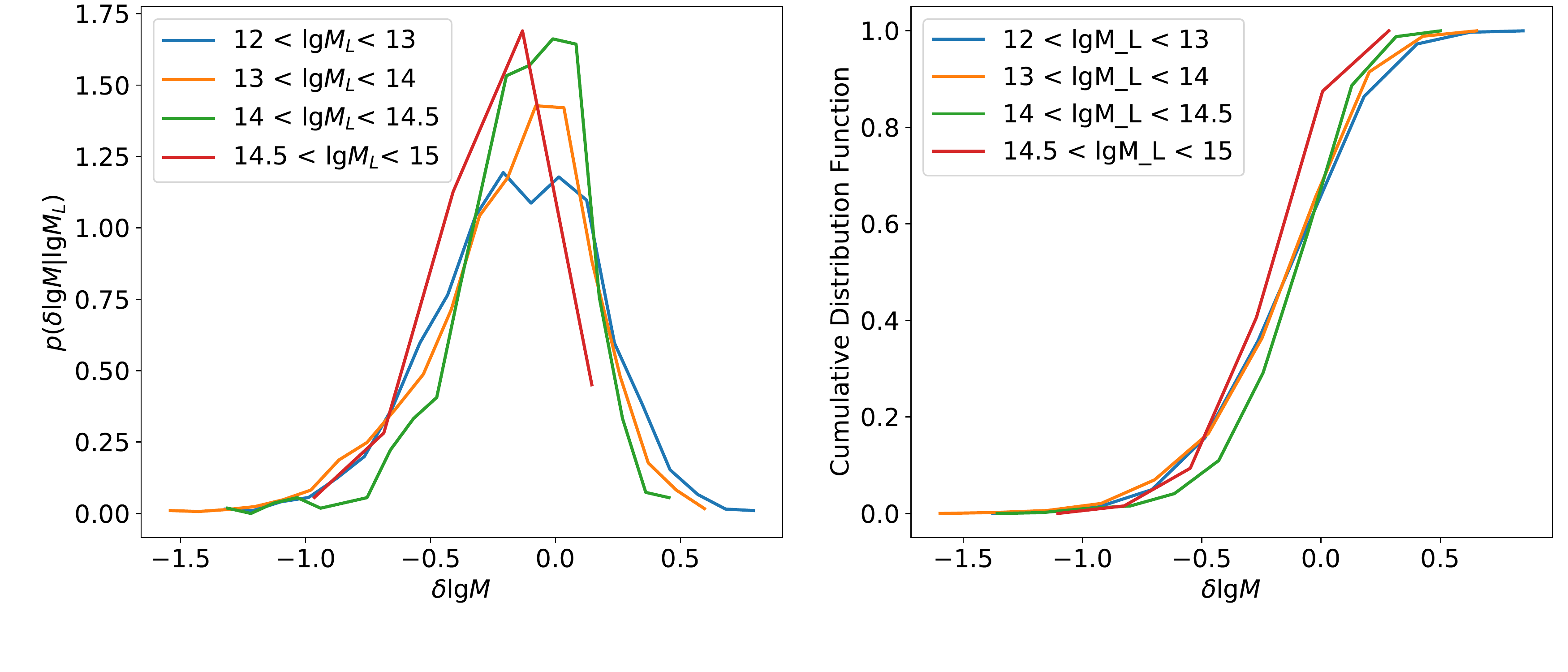}
 		\caption{The PDF (left) and CDF (right) of $\delta \lg M$ as a function of $\lg M_L$ in different redshift bins. In the mass bin $14.5 < \lg M_L < 15$, the number of cluster in the mock catalog is too less to construct a precise description of the PDF. This may induce bias into 1h-term template estimation. \label{fig:pdf}}
	\end{figure*}
	
	In DESI group catalog DR9, the assigned halo mass $M_L$ and halo true mass $M_t$ is not a one-to-one relation. In Fig.\ref{fig:mlmt}, we show the difference $\delta \lg M$ of $\lg M_L$ and $\lg M_t$ and its scatter $\sigma_{\lg M}$. The mean value and the scatter are all dependent on $\lg M_L$. The amplitude of 1h-term is proportional to $M^{5/3}$. Thus the uncertainty of 0.2 dex for mass means 2.2 times difference for the 1h-term profile. Therefore it is necessary to calibrate the cluster mass when calculating the 1h-term template. Otherwise the measurement $f_{\rm gas}$ would be catastrophically biased.  We show how to obtain an appropriate 1h-term profile according to the $\lg M_l- \lg M_t$ distribution relation as follows.\\
	
	Using these mock data, we can obtain the possibility distribution of $\delta\lg M$ in the mass bin $\lg M_1 <  \lg M <  \lg M_2$. Fig.\ref{fig:pdf} shows the PDF and CDF of $\delta \lg M$ in four redshift bins. 
	Then the 1h-profile of an observed cluster sample is 
	\ba
	\label{eq:yt}
		y_{\theta}=\int^{z_{\rm max}}_{z_{\rm min}} \int^{\lg M_{\rm L,max}}_{\lg M_{\rm L,min}} \int^{\lg M_t=\infty}_{\lg M_t=0} y_\theta(\lg M_t,\ z)n(\lg M_t,z)\nonumber\\ p(\delta \lg M= \lg M_t-\lg M_L|\lg M_L)d\lg M_td\lg M_Ldz,
	\ea
	$y_\theta(\lg M_t,\ z)$ is the y-profile of a halo whose mass and redshift are $M_t$ and $z$, $n(\lg M_t,z)$ is the number of clusters in the mass bin $(\lg M_t-\frac{d\lg M_t}{2}, \lg M_t+\frac{d\lg M_t}{2})$ and redshift bin $(z-\frac {dz} 2, z+\frac {dz} 2)$, and $p(\delta \lg M|\lg M_L)$ is the PDF of $\delta\lg M$ at $\lg M_L$. When the bin length $d\lg M_L$ and $dz$ small enough, Eq.\ref{eq:yt} is the unbiased 1h-term template for the cluster sample with $z_{\rm max} <  z <  z_{\rm min}$  and $\lg M_{\rm L,max}  <  \lg M_L  <   \lg M_{\rm L,min}$.\\

	Here we do not consider the uncertainty of the PDF $p(\delta \lg M|\lg M_L)$. However, the number of halos is decreasing with the halo mass, so there are only a few massive halos in massive mass bins. This may cause a large uncertainty on the estimate of the mean value of $\delta \lg M$ and the total PDF at large $\lg M_t$. In addition, \citet{wang2022} shows there is a slight difference between halo masses determined by their ESD model and provided by \citet{Yang2021} when $\lg M_{\rm L}\sim 14.8$. Therefore we abandon the most massive bin in redshift $0.2 < z < 0.4$ and $0.4 < z < 0.6$, whose $M_L \geq 10^{14.9}M_\odot/h$, in fitting relation of $f_{\rm gas}$ and halo mass $M$.
	
	In Sec.\ref{subsec:YM}, we obtain a sightly redshift-evolving Y-M relation. Here, we want to point out a possible redshift-dependent systematic error in mass estimation may influence this relation. In \citet{wang2022}, Fig.7 shows a redshift evolution of the difference between the cluster mass determined by cosmic shear and that given by Y21 catalog. However, when modifying the cluster mass with simulation, we do not take the redshift dependence into consideration due to the scarcity of simulation data.

\section{Concentration}\label{app:concentration}
	\begin{figure*}
		\centering
 		\includegraphics[width=\textwidth]{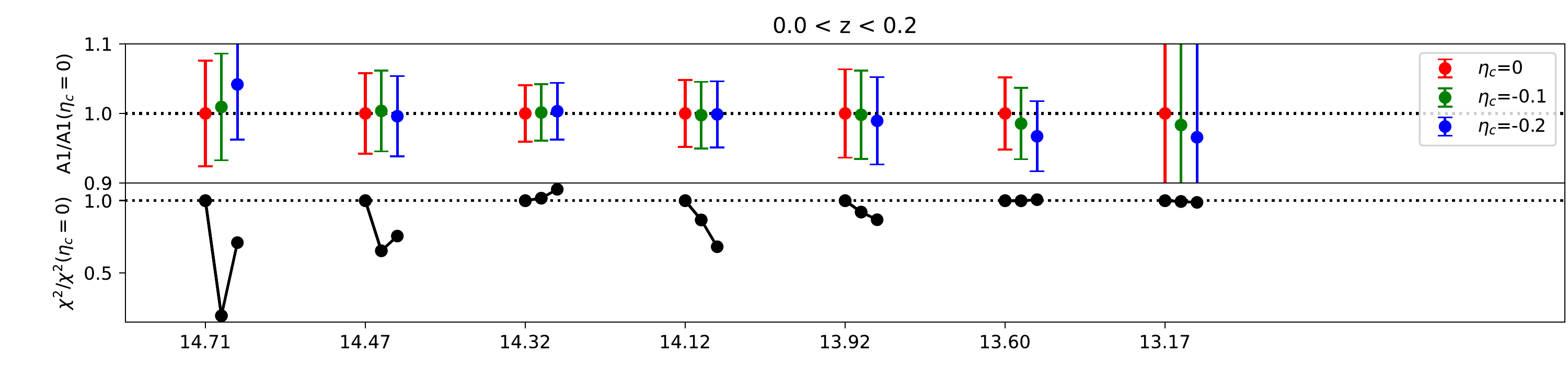}
 		\includegraphics[width=\textwidth]{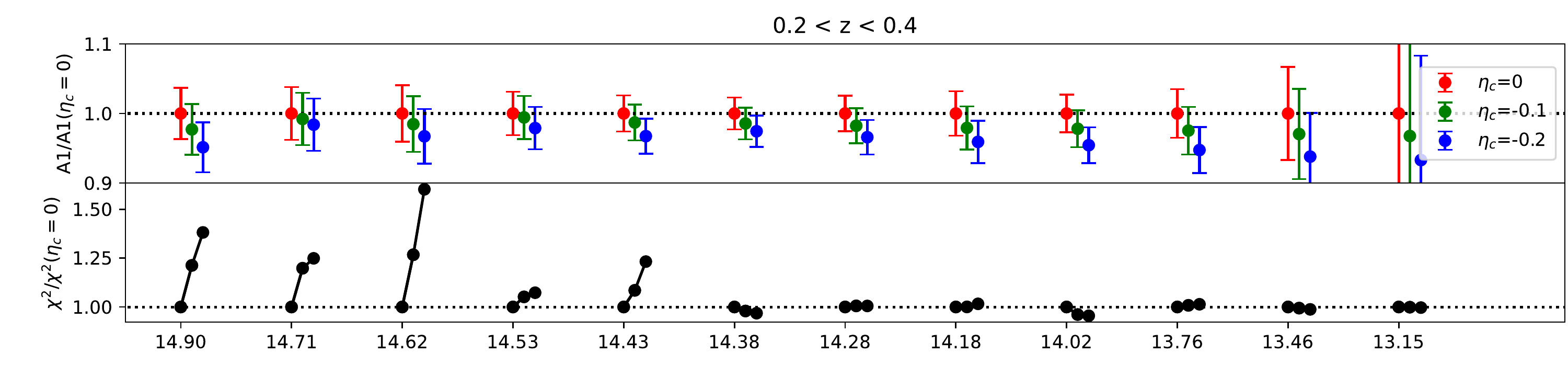}
 		\includegraphics[width=\textwidth]{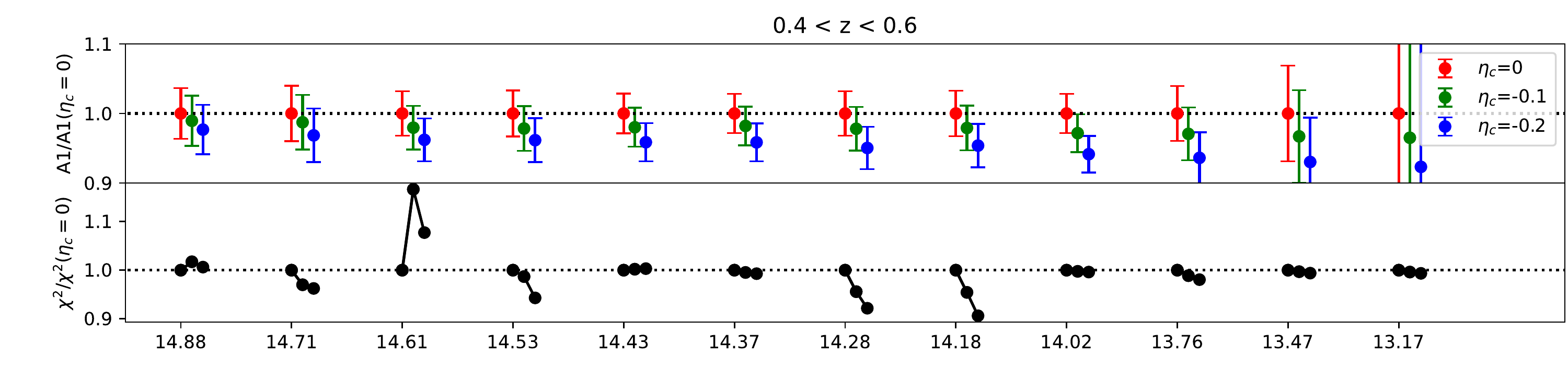}
 		\includegraphics[width=\textwidth]{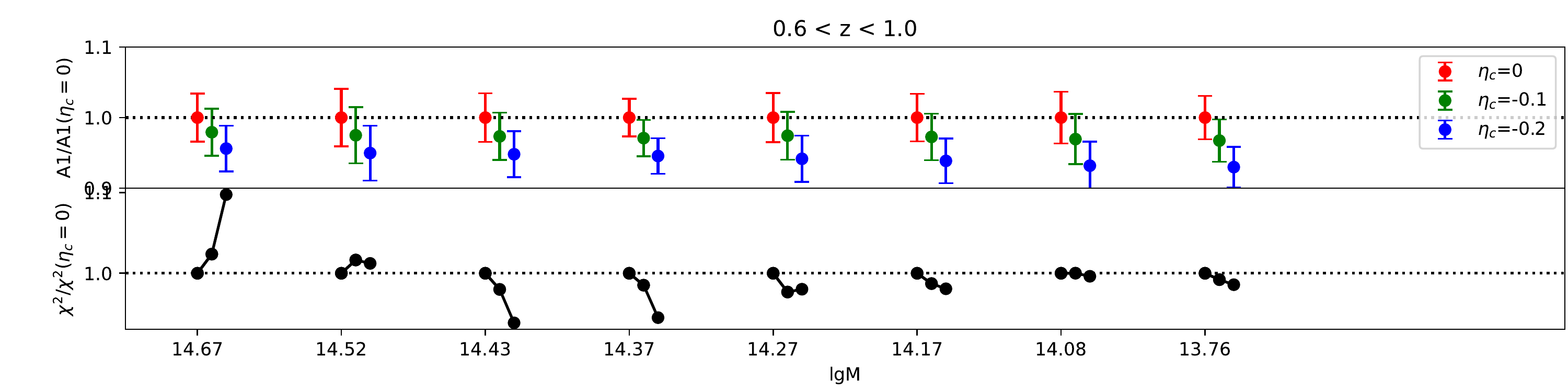}
 		\caption{The coefficient of the 1h-term influenced by concentration model. The three different concentration models are $\eta_{c}=0, -0.1, -0.2$ represented by red, green and blue points. $\eta_c$ is the parameter thar characterizes how the concentration is sensitive to the fraction of baryon in a halo (Eq.\ref{eq:c_modi}). The redshift bin is the label on the top of each panel. The top part of each panel shows $A_1$ of each mass bin for the three models. And to make the figure more readable, each point is normalized by the $A_1(\eta_c=0)$ in the same bin. The bottom part shows the $\chi^2_{\rm min}$. \label{fig:c}}
	\end{figure*}

	In the fiducial measurement, we assume the concentration-mass relation from \citep{Duffy2008} for dark matter halos. However, the effects from baryon would influence the concentration and the 1h-term profile. In Section.\ref{subsec:AR_1h}, it has been observe more than a half of baryon is blowed away from the halo with $M< 10^{14}M_\odot/h$. Therefore, these halos would become more loose. Follow \citep{Mead2020}, we adopt the method to modify the concentration by unbound gas
	\ba\label{eq:c_modi}
		c_{\rm new} (M)= c(M)\left[ 1+\eta_{c}\left(1-\frac{f_{\rm gas}}{\Omega_b/\Omega_m}\right) \right].
	\ea
	When $f_{\rm gas}=0$, $c_{\rm new} (M)= c(M)$ the same as the fiducial case. And when all of the baryon is blowed away from the halo, $f_{\rm gas}=1$, $c_{\rm new}(M)=(1+\eta_c)c(M)$. The factor $1+\eta_c$ can characterize how the concentration of a halo would change, if it lose all baryons.
	Here, we compare three situation with $\eta_c=0, -0.1, -0.2.$ In Fig.\ref{fig:c}, we show the comparison of these three situations for four redshift bins. For each redshift bin, the top panel shows the ratio of the fitted $f_{\rm gas}$ and the fiducial case. The bottom panel shows the ratio of $\chi^2_{\rm min}$ and the fiducial case. \\
	
	For massive, low-redshift cluster samples, the fitted $f_{\rm gas}$ increases with the decreasing of $\eta_c$. For other cluster bins, the fitted $f_{\rm gas}$ decreases with the decreasing of $\eta_c$. And the low-mass clusters are more sensitive to the modification of concentration because their baryon abundance is lower. However, the measurement uncertainties of the low-mass cluster sample are also larger.\\
	
	Fortunately, all differences cannot be distinguished by the 1-$\sigma$ errorbar. Therefore, the simple fiducial assumption of concentration would not bias the measurements.
	On the other hand, this means our measurement cannot raise a constraint on the halo concentration currently. 
	
	With more accurate measurements with the upcoming surveys, the concentration (or the shape of density profile) may become a non-negligible ingredient. Fortunately, the shear-group correlation could put a constraint on halo concentration as a function of redshift and halo mass \citep{wang2022}.
	
\section{Mis-centering}\label{app:miscentering}
	\begin{figure*}
		\centering
 		\includegraphics[width=\textwidth]{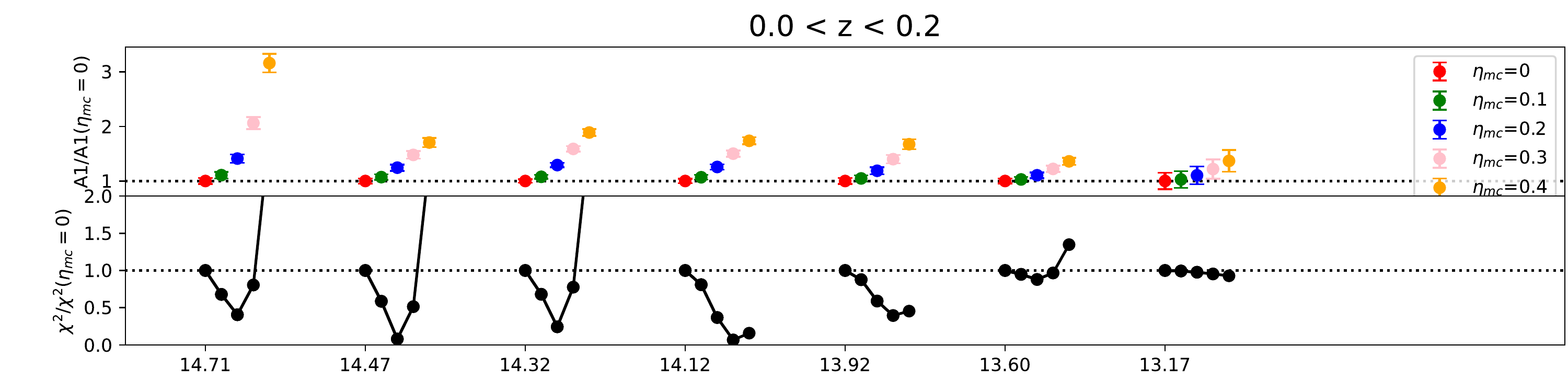}
 		\includegraphics[width=\textwidth]{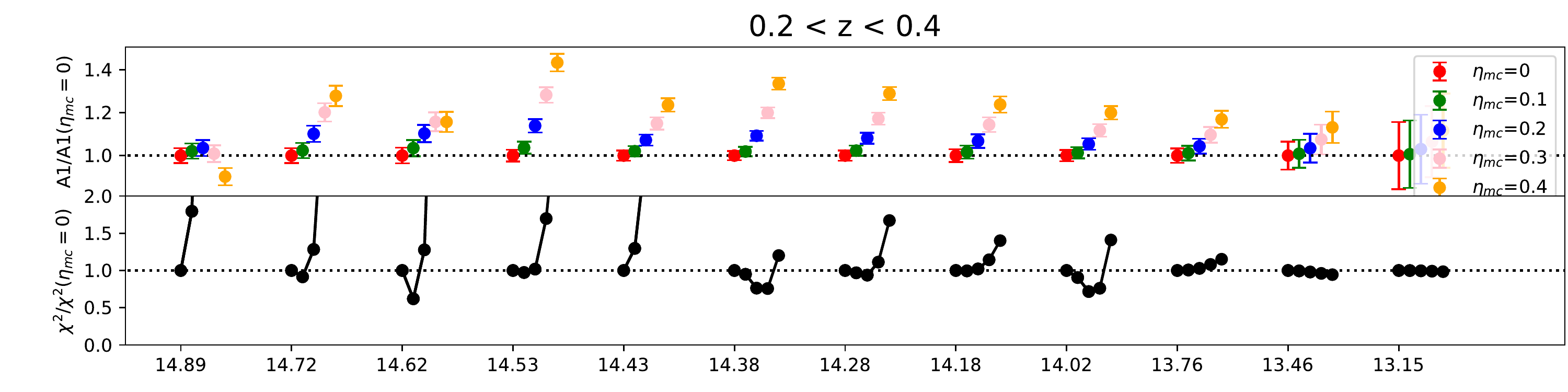}
 		\includegraphics[width=\textwidth]{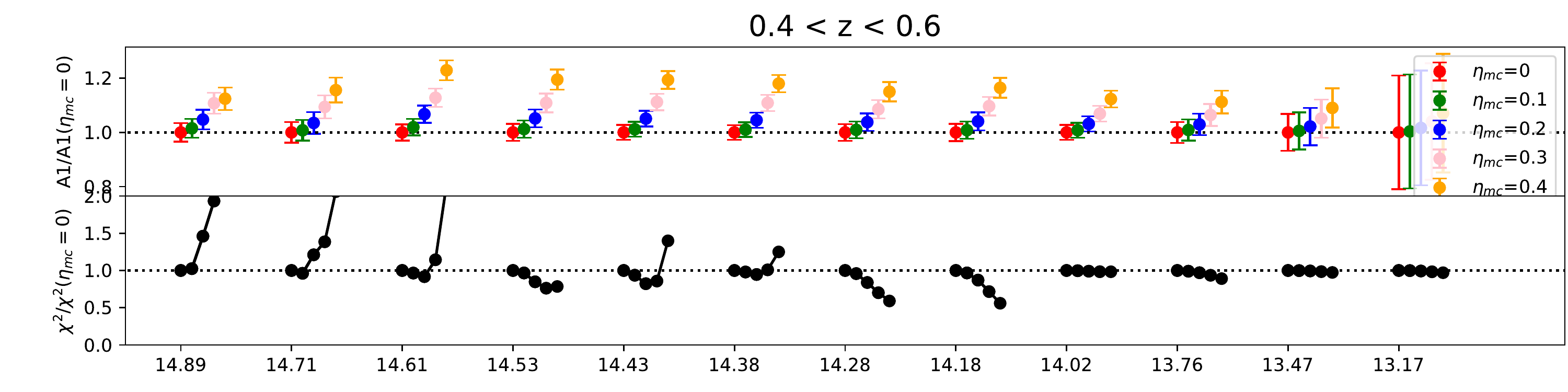}
 		\includegraphics[width=\textwidth]{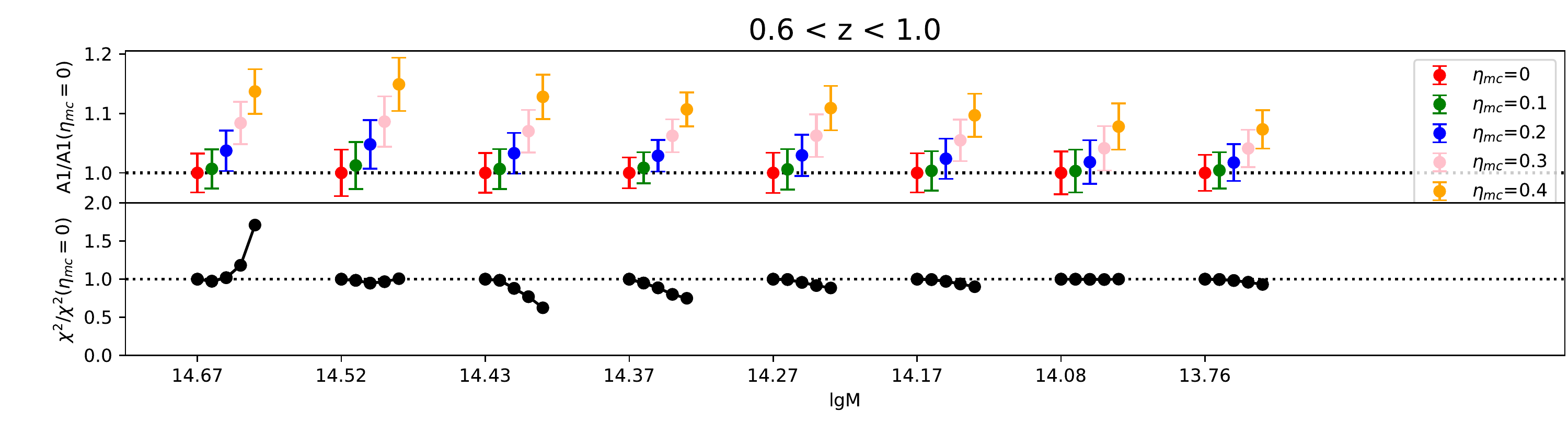}
 		\caption{The same with Fig.\ref{fig:c} but for different mis-centering models with $\eta_{\rm mc}=0, 0.1, 0.2, 0.3, 0.4$. \label{fig:mc}}
	\end{figure*}

	In DESI group catalog DR9, the position of clusters would have misalignments with the actual minimum gravitational potential points of them. This mis-centering would suppress stacked tSZ profiles. Here, we treat this effect as the same way of the beam in the CMB survey when generating 1h- and 2h-term templates. We assume the amplitude of the mis-centering is proportional to the virial radius of a halo.
	\ba
		\sigma_{\rm mc}=\eta_{\rm mc}\frac{r_\nu}{d_c}
	\ea
	And in the templates of 1h- and 2h-term, the parameter $\sigma_{\rm beam}$ is replaced by $\sigma_{\rm eff}=\sqrt{\sigma_{\rm beam}^2+\sigma_{\rm mc}^2}$.
	
	In Fig.\ref{fig:mc}, we test how the parameter $\eta_{\rm mc}$ would influence the measurements. In the top panel of each redshift, it shows the ratio of the $f_{\rm gas}$ with $\eta_{\rm mc}=0, 0.1, 0.2, 0.3, 0.4$ and case with $\eta_{\rm mc}=0$. And the bottom panels show the ratio of $\chi_{\rm min}^2$. \\
	
	We find the $\eta_{\rm mc}$ indeed have a large influence on the fitting, especially for the low redshift, massive cluster samples. For small clusters and high redshift clusters, the $\sigma_{\rm mc}$ is smaller than the beam size $\sigma_{\rm beam}$ , due to small $r_{\nu}$ or large $d_c$. In this situation, the fitting is less sensitive to the mis-centering effect. Low-redshift samples suffer from mis-centering effect more seriously. For most cluster samples with $0\leq z < 0.2$, when $\eta_{\rm mc}={0.2}$, $\chi^2_{\rm min}$ reaches the minimum point. Therefore, we set $\eta_{\rm mc}={0.2}$ in the fiducial measurement. 

\section{Low-mass group sample}\label{APP:low_mass_sample}
    \begin{deluxetable*}{cccccccccccc}
    \tablecaption{Detail infomation of low-mass groups measurements \label{tab:small_mass}}
	\tablehead{\colhead{range of $z$} & \colhead{range of $\lg M_{\rm L}$} & \colhead{$N_{\rm cluster}$} & \colhead{$\lg M_t$} & \colhead{$\bar b_g$} & \colhead{$A_1$} & \colhead{S/N($A_1$)} & \colhead{$A_2$} & \colhead{S/N($A_2$)} & \colhead{$A_3\times 10^8$} & \colhead{S/N($A_3$)} & \colhead{$\chi^2_{\rm min}$} 
	}
	\startdata
	$[$0.0, 0.2) & $[$11, 13) & 66648 & 12.69 & 1.14 & 2.311 & 3.65 & 0.51 & 10.9 & -1.36 & 4.0 & 14.81 \\
    $[$0.2, 0.4) & $[$11, 13) & 102191 & 12.8 & 1.26 & 5.164 & 5.92 & 0.61 & 13.1 & -0.81 & 4.9 & 12.866 \\
    $[$0.4, 0.6) & $[$11, 13) & 29076 & 12.86 & 1.4 & 5.09 & 2.7 & 0.59 & 3.9 & 1.22 & 4.2 & 10.013 \\
    $[$0.6, 1.0) & $[$13, 13.5) & 20361 & 13.34 & 2.07 & -0.309 & 0.63 & 0.5 & 1.7 & 0.58 & 1.7 & 2.977 \\
    \enddata
    \end{deluxetable*}
    
    \begin{figure}
		\centering
 		\includegraphics[width=0.5\textwidth]{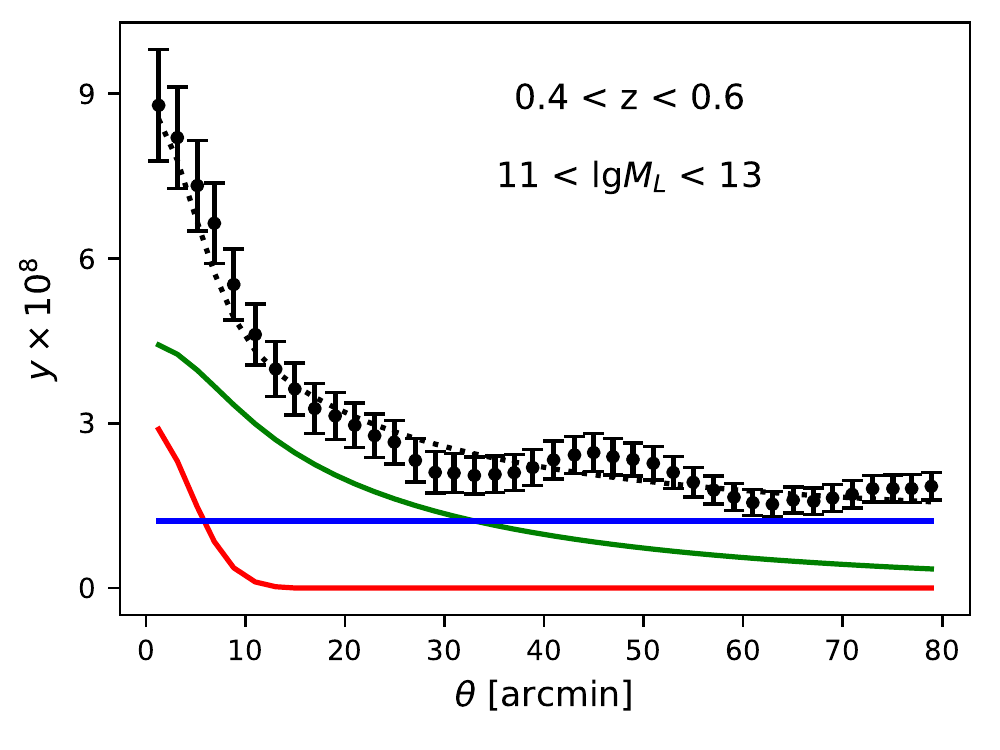}
 		\caption{The stacking results of tSZ measurements as a function of angular distance to the center of clusters for a sample with $0.4 \leq z < 0.6$ and $11 \leq \lg M < 13$. The black dots are the stacking results with errorbars estimated by Jackknife resampling. The red, green and blue lines represent the bestfit one-halo, 2h-and the background term. The dotted lines are the sum of these three terms.\label{fig:small_mass}}
	\end{figure}
    In this suction, we show the tSZ measurement of group with $\lg M<13$ ($\lg M<13.5$ for $0.6 \leq z < 1$). An example of $0.4 \leq z < 0.6$ is shown in Fig.\ref{fig:small_mass}. Results of other redshift bins are shown in Fig.\ref{tab:small_mass}. At these mass range, the amplitude of the background is comparable to that of the 1h- and 2h-terms. And the amplitude of 1h-term becomes unreasonable. This means the shape of the background plays an important role in the fitting results. And the large fluctuations at $\theta > 20$ arcmin may indicate a scale-dependent background. Therefore we do not include these results in main body analysis.

\bibliography{stack_tsz}{}
\bibliographystyle{aasjournal}

\end{document}